\documentclass[PAPER, atlasdraft=false, coverpage=false, cernpreprint, USenglish, texmf, orcidlogo]{atlasdoc}
\usepackage{atlaspackage}
\usepackage{atlasbiblatex}

\usepackage{atlasphysics}
\usepackage{atlasheavyion}
\graphicspath{{logos/}{figures/}}

\usepackage{ANA-HION-2026-01-PAPER-defs}

\AtlasTitle{Measurement of jet photoproduction in ultra-peripheral \ce{Pb}+\ce{Pb} collisions
without nuclear breakup at $\sqrt{s_\mathrm{NN}} = 5.02$ TeV with the ATLAS detector}

\AtlasAbstract{%
In ultra-relativistic heavy ion collisions at the LHC, each nucleus
acts as a source of high-energy quasi-real photons that can participate in
scattering processes without causing either participating nucleus to
break up and emit forward neutrons. This paper extends recent measurements
of $\gamma+A\rightarrow\mathrm{jets}$ production in ultra-peripheral \ce{Pb}+\ce{Pb} collisions
at $\sqrt{s_\mathrm{NN}} = 5.02$~TeV with forward neutron emission on exactly
one side of the event. The data presented here was
recorded by the ATLAS collaboration at the LHC in 2018, corresponding to a
luminosity of $1.72$~nb$^{-1}$. These results examines
$5.02$~TeV \ce{Pb}+\ce{Pb} collisions where neither nucleus breaks up ($0n0n$), providing a mixture of
photon--pomeron ($\gamma+I\!\!P\rightarrow\mathrm{jets}$),
photon--photon ($\gamma+\gamma\rightarrow\mathrm{jets}$), and peripheral
photonuclear ($\gamma+A\rightarrow\mathrm{jets}$) events. The different processes
are statistically separated via a template fit of the minimum rapidity
gap distribution. The kinematics of the hard processes are determined
from $R = 0.4$ jets reconstructed using the anti-$k_t$ algorithm. The
statistical separation of the different processes then allows for the
first measurement of $\gamma+I\!\!P\rightarrow\mathrm{jets}$ cross-sections in nuclear collisions at the LHC. The
rate for electromagnetic dissociation of $0n0n$ $\gamma+A\rightarrow\mathrm{jets}$ events is
also measured and compared to the analogous result from collisions with
single-sided neutron emission. These comparisons support the hypothesis that
$\gamma+A\rightarrow\mathrm{jets}$ events without forward neutron emission select
a more peripheral class of $\gamma+A$ collisions.

}

\AtlasRefCode{HION-2026-01}

\PreprintIdNumber{CERN-EP-2026-125}

\AtlasJournal{Phys.\ Rev.\ C}
\AtlasCoverEgroupAnalysisTeam{atlas-HION-2024-12-analysis-team@cern.ch}

\hypersetup{pdftitle={ATLAS document},pdfauthor={The ATLAS Collaboration}}

\begin{document}

\maketitle

\section{Introduction}
\label{sec:intro}

In ultra-relativistic heavy ion collisions, the
electromagnetic fields associated with the highly charged ions
provide an intense flux of photons that make possible the study of
photon-induced scattering processes~\cite{Bertulani:2005ru,Klein:2020fmr}.
At two-nucleus impact parameters ($b_\text{AA}$) larger than
the sum of the nuclear radii, hadronic interactions between the nuclei
are strongly suppressed, and the photon-induced processes, commonly
referred to as ultra-peripheral collisions (UPCs),
dominate. UPCs in which
the photon emitted by one nucleus scatters off the opposing nucleus,
photonuclear (\gamA) collisions, probe the structure
of that nucleus, particularly at larger momentum
transfer~\cite{HION-2022-15,CMS-HIN-24-003}. Then, measurements of
the scattering cross-section may
provide direct sensitivity to the nuclear parton
distribution functions (PDFs)~\cite{Helenius:2018bai,Guzey:2019kik}.
Often, such hard processes proceed via color exchange with the
struck nucleus, and the resulting particle production typically fills
the rapidity region between that nucleus and the outgoing products of
the hard scattering. The hard scattering process has a high
probability of exciting the struck nucleus, causing subsequent emission of ``forward''
neutrons with rapidity approximately equal to that of the
nucleus which emits
them ($|\eta| > 8$)~\cite{Strikman:1998cc,Larionov:2018igy,Alvioli:2024cmd}.
However, \gamA hard-scattering processes can also proceed via
no net color exchange with the nucleus. Such ``diffractive'' processes
are dominated by coherent pomeron ($I\!\!P$) emission from the struck nucleus.
Coherent \gamA diffractive processes producing exclusive final states,
especially vector mesons, have been
extensively studied~\cite{CMS-HIN-22-002,ALICE:2024whv,STAR:2023nos,HION-2025-01},
as they probe nuclear parton densities at low
Bjorken-$x$, and may be sensitive to the effects of gluon saturation.
Non-exclusive final states in diffractive \gamA scattering have
received less attention experimentally, but they may provide another
valuable avenue for studying the low-$x$ structure of the nucleus~\cite{Guzey:2016tek,Iancu:2023lel}.

\begin{figure}[!t]
\centerline{
\subfloat[]{\includegraphics[width=0.315\textwidth]{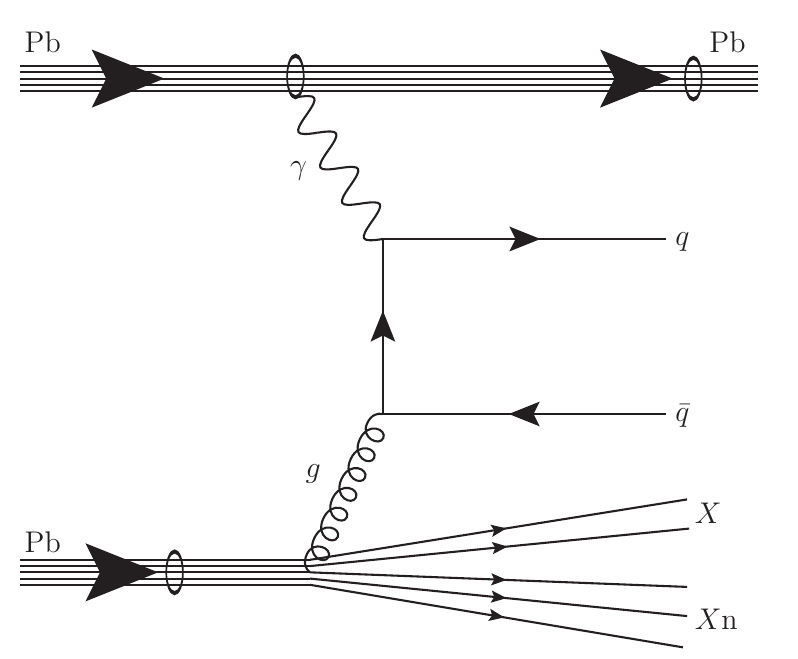}}
\subfloat[]{\includegraphics[width=0.345\textwidth]{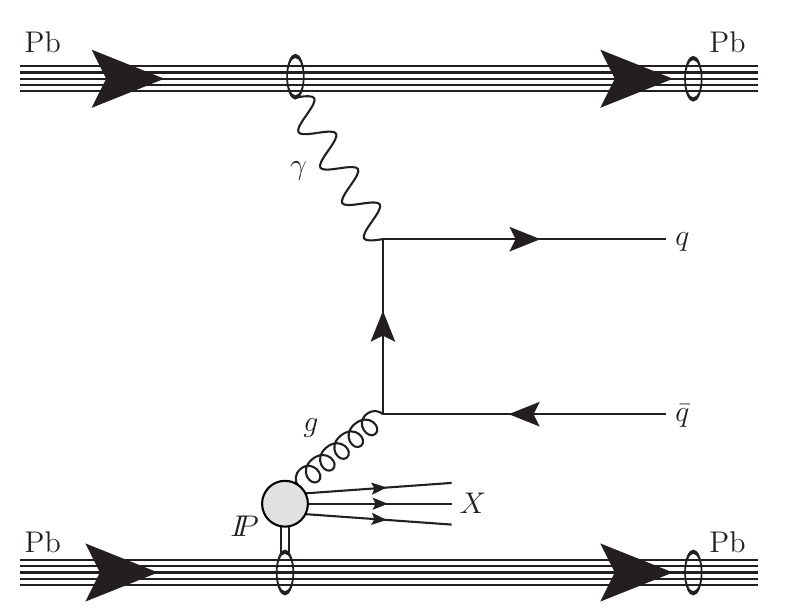}}
}
\centerline{
\subfloat[]{\includegraphics[width=0.33\textwidth]{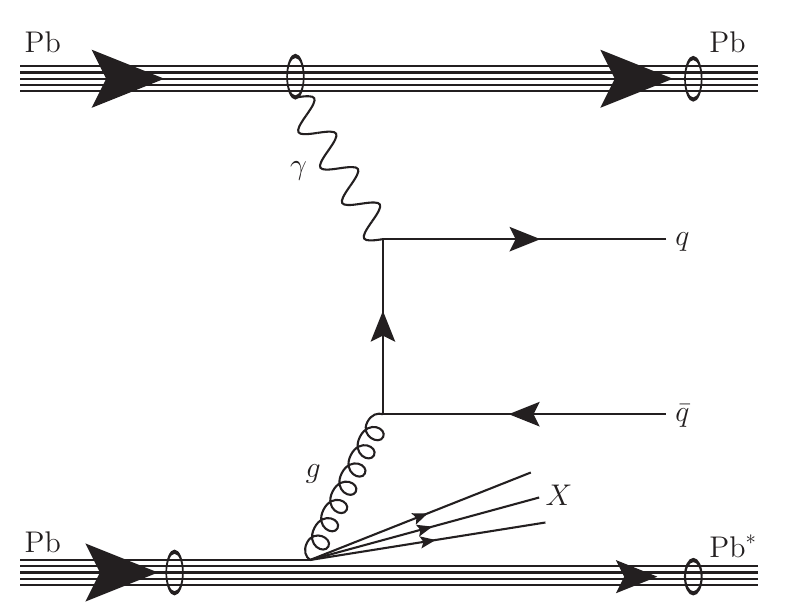}}
\subfloat[]{\includegraphics[width=0.33\textwidth]{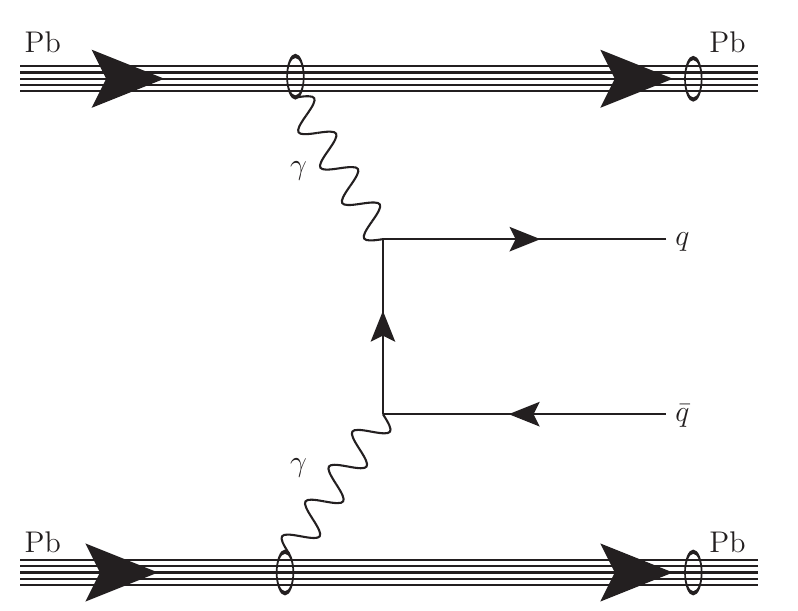}}
}
\caption{
Example Feynman diagrams describing the direct photon contribution to
four UPC jet production processes, three of which (b-d) contribute to \znzn\ UPC jet production. The resolved photon processes contribute
to this measurement but are not shown here.
(a) A non-diffractive \gamA interaction that induces neutron emission in
the direction of the struck nucleus, (b) A diffractive interaction proceeding
through \gamPom\ scattering which
produces a pair of jets, leaving a pomeron fragment to partially fill the rapidity
gap between the jets and the pomeron-emitting nucleus. (c) \gamA\ interaction
where the photon strikes a nucleon near the edge of the nucleus, producing a pair
of jets while leaving the nucleus intact. (d) A \gamgam\ interaction
where both nuclei emit photons which produce a pair of jets.
}
\label{fig:feynman}
\end{figure}

Photonuclear collisions producing two or more jets (see diagrams in
Figure~\ref{fig:feynman}) have long
been considered a valuable tool for probing nuclear parton
distributions~\cite{Strikman:2005yv,Helenius:2018bai,Guzey:2019kik},
especially as the kinematics of the outgoing jets
provide a direct measurement of the longitudinal momenta of the
particles entering the hard scattering.
Indeed, a recent measurement by ATLAS~\cite{HION-2022-15} of the
non-diffractive \gamAjets\ (Figure~\ref{fig:feynman}a) cross-section in
$\sqn = 5.02$~TeV \PbPb\ collisions at the LHC is sensitive to the nuclear
PDF (nPDF) in lead with precision better than 5\% over a wide range of momentum
fraction (\xA) and momentum transfer ($Q^2$) that has only limited
constraints from other measurements. Diffractive jet production (Figure~\ref{fig:feynman}b), which
was studied extensively in $e+p$ collisions at
HERA~\cite{ZEUS:2007uvk,H1:2010xdi,H1:2012xlc,H1:2015okx}, is of
great interest in nuclear collisions as it provides sensitivity to previously
unmeasured nuclear diffractive PDFs~\cite{Guzey:2016tek}, which are
expected to be strongly suppressed by nuclear shadowing effects~\cite{Guzey:2020ehb}.
Also, photo-diffractive measurements involving nuclei may provide insight
on poorly understood factorization-breaking~\cite{Guzey:2016tek,Klasen:2004qr} that
may be enhanced by the nuclear target~\cite{Guzey:2020ehb}.
Diffractive jet production will be a subject of intensive study in
deep-inelastic scattering at the future Electron Ion Collider
(EIC)~\cite{Iancu:2022lol}, and measurements in UPC processes will both
complement and lay the groundwork for those future measurements.

Diffractive and non-diffractive \gamAjets processes can be
distinguished~\cite{HION-2022-15} through a combination of neutron
topology and rapidity gaps -- regions in pseudorapidity with no hadron production. As
noted above, non-diffractive \gamA collisions have a high
probability for stimulating neutron emission from the struck nucleus (a \znxn\ topology),
while coherent diffraction has a high probability for leaving the
nucleus intact~\cite{Strikman:1998cc,Larionov:2018igy,Alvioli:2024cmd}.
Thus, the measurement in Ref.~\cite{HION-2022-15} suppressed
diffractive processes by requiring at least one neutron to be emitted in the direction of the
struck nucleus. By contrast, a measurement that requires {\em no} neutrons in the
direction of the struck nucleus will strongly suppress the
non-diffractive \gamAjets contribution, thereby allowing a
measurement of the diffractive cross-section. It is possible,
however, that non-diffractive \gamA scatterings may occur on the
periphery of the nucleus and leave the nucleus
intact~\cite{Larionov:2018igy,Alvioli:2024cmd}
(Figure~\ref{fig:feynman}c), providing a background to
coherent diffraction. Such processes have been studied in greater detail
in Ref.~\cite{OtherPaperPRL}, as they may provide unique sensitivity
to the impact parameter dependence of the nPDF. A notable feature of
non-diffractive \gamA processes is that there are typically large gaps
in the direction of the photon-emitting nucleus and small or no gaps
in the direction of the struck nucleus. In contrast, diffractive \gamA\ processes
are expected to exhibit rapidity gaps in both directions~\cite{Guzey:2016tek,Guzey:2019dpp},
with the larger gap typically in the direction of the photon-emitting nucleus. Thus, an
analysis of rapidity gaps can help distinguish diffractive \gamAjets\
events from peripheral non-diffractive events.

Photonuclear measurements are often performed by requiring no neutrons
in the direction of the photon-emitting nucleus. This requirement
takes advantage of the fact that the photon flux is dominated
by coherent emission that leaves the source nucleus
intact~\cite{Klein:2016yzr,Klein:2020fmr}. Then, \gamAjets\ events with no forward neutrons in either
direction -- so-called \znzn events -- are expected to be dominated by
a combination of diffractive and peripheral non-diffractive
scattering. A background to these measurements arises from
two-photon (\gamgam) scattering processes (Figure~\ref{fig:feynman}d)
that may also produce jets
and have a high probability of leaving both nuclei
intact. Thus, \gamAjets\ production with a \znzn\ topology would have contributions
from three processes whose Feynman diagrams are shown in
Figure~\ref{fig:feynman}(b-d). These processes can be distinguished by
their different typical jet kinematics and by the different behavior of the
rapidity gaps.

A well-known issue in using neutron topology to study
UPCs is that excitation of one or both nuclei by the long-range
Coulomb fields of the nuclei -- electromagnetic dissociation (EMD) --
can stimulate additional neutron emission, altering the forward
neutron topology~\cite{PhysRevLett.109.252302,ALICE:2022iqi}.
In the case of a \znzn\ measurement, events with
extra EMD neutrons would be excluded. The probability
for EMD emission of neutrons depends on
$b_\text{AA}$~\cite{HION-2023-13,CMS-HIN-19-014,CMS-HIN-22-002,Guzey:2013jaa},
so calculations of the EMD rate (such as in Ref.~\cite{Eskola:2024fhf})
necessarily depend on assumptions regarding the impact parameter
distribution. A separate measurement of ``\xnzn'' \gamAjets with
emission of neutrons in the direction consistent with the large,
photon-associated gap -- $\znzn + \text{EMD}$ processes -- is
particularly helpful in constraining such calculations.
In addition, a measurement of the EMD rate
in {\em non-diffractive} \gamAjets events without neutron
emission from the hard-scattering process provides
sensitivity to the distribution of photon impact
parameters with respect to the struck nuclei. If the measured
EMD rate is smaller in \znzn\ collisions than in \znxn, that would
provide indirect confirmation that the non-diffractive \znzn\
processes are truly ``peripheral''.

This paper presents a measurement by ATLAS in
\PbPb\ UPCs at $\sqn = 5.02$~\TeV  of \gamAjets\ and \gamPomJets\ scatterings that have no
forward neutrons, and have, thus, a \znzn\ topology. A template-fitting
procedure is implemented to compute signal fractions
for the different physical processes shown in Figure~\ref{fig:feynman}
that contribute to \znzn\ jet production.
Using this procedure to statistically separate the different components,
the \gamPomJets\ cross-section is measured double-differentially in the jet scalar sum \pT\ and
jet system rapidity. The integrated cross-sections are compared to theoretical predictions corresponding
to different scenarios for the size of factorization-breaking effects. Additionally, the
rate for EMD emission of neutrons in what would otherwise be \znzn\ \gamA\ collisions
is measured and compared to the corresponding rate for \znxn\ \gamA\ collisions.

The remainder of this paper is organized as follows:
Section~\ref{sec:detector} describes the ATLAS detector; Section~\ref{sec:samples}
lists the relevant data and simulated event samples used in these results;
Section~\ref{sec:analysis} describes the reconstruction, event selection, and
data analysis methodology; Section~\ref{sec:emd} discusses the techniques used to
perform measurements of electromagnetic dissociation processes; Section~\ref{sec:systematics}
describes the systematic uncertainties; Section~\ref{sec:results} presents the results;
and Section~\ref{sec:conclusion} draws conclusions from the presented measurements.


%

%
\section{The ATLAS detector}
\label{sec:detector}


The ATLAS detector~\cite{PERF-2007-01} at the LHC covers nearly the entire solid angle around the collision point.\footnote{ATLAS uses a right-handed coordinate system with its origin at the nominal interaction point (IP)
in the center of the detector and the \(z\)-axis along the beam pipe.
The \(x\)-axis points from the IP to the center of the LHC ring,
and the \(y\)-axis points upwards.
Polar coordinates \((r,\phi)\) are used in the transverse plane,
\(\phi\) being the azimuthal angle around the \(z\)-axis.
The pseudorapidity is defined in terms of the polar angle \(\theta\) as \(\eta = -\ln \tan(\theta/2)\) and is equal to the rapidity
$ y = \frac{1}{2} \ln \left( \frac{E + p_z}{E - p_z} \right) $ in the relativistic limit.
Angular distance is measured in units of \(\Delta R \equiv \sqrt{(\Delta y)^{2} + (\Delta\phi)^{2}}\).}
It consists of an inner tracking detector surrounded by a thin superconducting solenoid, electromagnetic and hadronic calorimeters,
and a muon spectrometer incorporating three large superconducting air-core toroidal magnets.

The inner-detector system (ID) is immersed in a \qty{2}{\tesla} axial magnetic field
and provides charged-particle tracking in the range \(|\eta| < 2.5\).
The high-granularity silicon pixel detector covers the vertex region and typically provides four measurements per track,
the first hit generally being in the insertable B-layer (IBL) installed before Run~2~\cite{ATLAS-TDR-19,PIX-2018-001}.
It is followed by the SemiConductor Tracker (SCT), which usually provides eight measurements per track.
These silicon detectors are complemented by the transition radiation tracker (TRT),
which enables radially extended track reconstruction up to \(|\eta| = 2.0\).
The TRT also provides electron identification information
based on the fraction of hits (typically 30 in total) above a higher energy-deposit threshold corresponding to transition radiation.

The calorimeter system covers the pseudorapidity range \(|\eta| < 4.9\).
Within the region \(|\eta|< 3.2\), electromagnetic calorimetry is provided by barrel and
endcap high-granularity lead/liquid-argon (LAr) calorimeters,
with an additional thin LAr presampler covering \(|\eta| < 1.8\)
to correct for energy loss in material upstream of the calorimeters.
Hadronic calorimetry is provided by the steel/scintillator-tile calorimeter,
segmented into three barrel structures within \(|\eta| < 1.7\), and two copper/LAr hadronic endcap calorimeters.
The solid angle coverage is completed with forward copper/LAr and tungsten/LAr calorimeter modules
optimized for electromagnetic and hadronic energy measurements, respectively.

The luminosity is measured mainly by the LUCID--2~\cite{LUCID2}
detector that records Cherenkov light produced by charged particles
passing through the quartz windows
of photomultipliers located close to the beampipe.  Two zero-degree
calorimeters (ZDCs)~\cite{Avoni:2025lmu}, located symmetrically at a
distance of $\pm 140$~m from the IP, measure neutral particles having
$|\eta|>8.3$. These are primarily neutrons emitted at small rapidity
separation from the incident nuclei. During Run~2 the ZDCs are
tungsten-quartz sampling calorimeters with a total of 4.4 interaction
lengths of absorber.

ATLAS uses a two-level trigger system~\cite{TRIG-2016-01} with the first-level (L1)
trigger system implemented in custom hardware,
and a software-based ``high-level trigger'' (HLT) running algorithms
similar to those used in offline analysis.

A software suite~\cite{ATL-SOFT-PUB-2021-001} is used in data simulation, in the reconstruction
and analysis of real and simulated data, in detector operations, and in the trigger and data acquisition
systems of the experiment.


%

%
%

%
\section{Data and simulated event samples}
\label{sec:samples}

This paper describes a series of measurements performed using $1.72$~nb$^{-1}$ of
\PbPb\ collision data at $\sqn = 5.02$~\TeV\ recorded in 2018 with the ATLAS detector. During the data-taking, the typical
number of hadronic interactions per bunch crossing was less than $0.005$,
while the rate for additional electromagnetic interactions in the bunch-crossing of interest (EMD pile-up) is larger
\cite{PhysRevLett.109.252302,PhysRevSTAB.17.021006,Klusek-Gawenda:2013ema},
causing forward neutron emission from additional dissociative \PbPb\ collisions
in up to $13\%$ of bunch crossings throughout the data-taking period.

The primary triggers used for the measurement impose requirements at L1
on the detected energy in the central calorimeter and in the ZDCs in order to
select \znzn UPC events
while vetoing hadronic \PbPb\ collisions. In particular, the L1
triggers required the total
transverse energy in the calorimeter, \sumET, to satisfy $5 < \sumET <
200$~GeV, and they rejected events having energy deposits
consistent with one or more neutrons in either ZDC.
The triggers also required one jet above a given transverse momentum
(\pT) threshold in the HLT. The HLT jet triggers are based on the anti-\kt
algorithm~\cite{Cacciari:2008gp,Fastjet} with radius $R = 0.4$  applied to topological clusters~\cite{PERF-2014-07}
formed from energy deposits in the calorimeter. The lowest nominal \pT
trigger threshold used in this measurement was 10~GeV, and it was fully %
efficient for single jets at 15~GeV (20~GeV) for $|\eta| > 3.2$
($|\eta|<3.2$). The  efficiency for the HLT to fire on any
of the jets in the final state is greater than 98\% for all events within
the fiducial acceptance of the measurement. For 40\%
of the sampled luminosity, the jet triggers were applied only over
the pseudorapidity range $\etajet < 3.2$, while for the remaining data,
the full range ($\etajet < 4.9$) was used.
In order to measure the EMD rate of \znzn\ \gamAjets\ collisions, an
additional sample was collected using the same jet trigger requirements but with
an L1 ZDC trigger requiring at least one neutron on exactly one side.

Several Monte Carlo (MC) samples were produced for this analysis using
the \pythia\ event generator~\cite{Bierlich:2022pfr,Helenius:2019gbd} for the three relevant
physical processes: \gamAjets, \gamPomJets, and \gamGamJets.
Final-state stable particles, defined as those with $c\tau > 10$~mm, were then
passed to a \textsc{Geant4}-based simulation of the ATLAS detector
\cite{Agostinelli:2002hh,SOFT-2010-01}, the output of which was
reconstructed in the same way as data.
Equal numbers of events were generated with photons propagating
in the positive and negative $z$ directions for both the
\gamA\ and \gamPom\ samples.
For each of the three samples,
the photon flux is modified in order to match the impact parameter distribution
calculated by \textsc{Starlight}~\cite{Klein:2016yzr}, and the simulated signal events include
both direct and resolved photon processes. The resolved photon case, where the photon interacts by fluctuating to a hadronic state, requires additional
modeling using the CJKL photon PDF set~\cite{Cor03}.

A large sample of \gamAjets\ events was produced using the methodology
outlined in Ref.~\cite{HION-2022-15} using the nCTEQ15~\cite{Kovarik:2015cma}
nuclear parton distribution functions and the A14 set of tuned
parameters ~\cite{ATL-PHYS-PUB-2014-021}. A sample of \gamGamJets was produced using the default
\pythia\ configuration with a photon flux corresponding to coherent emission
by a lead nucleus.

The ability to simulate hard diffraction using a configurable pomeron flux was added in
\PYTHIA[8.308] in order to simulate \gamPomJets\ processes
with coherent nuclear pomeron emission. The pomeron flux is described using the
Streng-Berger parameterization
~\cite{BERGER1987704,Streng:185039}:
\begin{equation}
\label{eqn:PomFlux}
x_{I\!\!P}f(x_{I\!\!P}, t) = A_{I\!\!P}\frac{e^{Bt}}{x_{I\!\!P}^{2\epsilon+2\alpha't}}.
\end{equation}
Here, $f(x_{I\!\!P}, t)$ is the pomeron flux, $x_{I\!\!P}$ is the pomeron energy
fraction relative to the emitting nucleus, and $t$ is the negative momentum-transfer
squared. The $B$ parameter typically characterizes the transverse size
of the pomeron emitter. For the proton, fits to diffractive deep-inelastic scattering
from H1~\cite{H1:2006uea,H1:2006zyl} yield parameters $B = 5.5~\text{GeV}^{-2}$,
$\alpha' = 0.06$~\GeV$^{-2}$, and $\epsilon = 1.118$.
For this analysis, \pythia was run with $\epsilon$ and $\alpha'$
unmodified while using a nominal $B$ parameter, $B_\mathrm{nom} = 200~\text{GeV}^{-2}$.
This value was derived by matching the pomeron $t$ distribution to the
nuclear form factor of lead. This modification to the pomeron flux also softens
the $x_{I\!\!P}$ distribution, such that the typical $I\!\!P$ and photon energies
in \gamPom\ interactions are very similar, consistent with the expection for both
being coherently emitted by a lead nucleus.


%

%
\section{Analysis}
\label{sec:analysis}

\subsection{Event reconstruction and selection}

Charged particle tracks in the ID are reconstructed over $|\eta^{\text{tr}}|<2.5$ using
the same methodology applied in minimum-bias proton--proton measurements~\cite{STDM-2015-02}.
Topological clusters (topo-clusters) are reconstructed in the pseudorapidity region
$|\eta^{\text{cl}}|<4.9$ from energy deposits in calorimeter cells~\cite{PERF-2014-07}.
Particle-Flow jets~\cite{PERF-2015-09} are reconstructed from combined calorimeter
and tracking information using
the anti-$k_t$ algorithm~\cite{Cacciari:2008gp} with radius parameter $R = 0.4$ as implemented in
the \verb+FastJet+~\cite{Fastjet} package. A dedicated jet calibration,
derived for detector conditions in UPC~\cite{HION-2022-15} collisions
using methods described in Ref.~\cite{JETM-2018-05}, is applied,

Offline requirements are applied to select events recorded during stable
running conditions of the LHC that have no detector hardware or readout
errors and that are not consistent with beam-induced backgrounds~\cite{HION-2022-15}.
UPC jet production events are selected from the triggered sample using a combination
of offline jet, ZDC, and
rapidity gap requirements. In particular, every event must have at least two reconstructed jets
satisfying the fiducial requirements: $\pTjet > 15$~GeV and $\etajet < 4.4$.
From all jets passing these requirements in a given event, the total scalar
transverse momentum (\HT), rapidity (\yjets), and mass (\mjets) are computed as:

\begin{equation}
\HT \equiv \sum_i p_{\mathrm{T}\,i} ,
\end{equation}
\begin{align}
\mjets & \equiv\left[\left(\sum_i E_{i}\right)^2-\left|\sum_i
\vec{p}_{i}\right|^2\right]^{1/2},\\
\yjets & \equiv\frac{1}{2} \ln{\left(\dfrac{ \sum_i E_{i} + \sum_i
{p_{z}}_i}{\sum_i E_{i} - \sum_i
{p_{z}}_i}\right)},\,
\end{align}
where $i$ runs over the $N$ measured jets in the event that satisfy the fiducial
requirements, $E$ and $\vec{p}$
represent jet energies and momentum vectors, respectively, and $p_{z}$
represents the longitudinal component of each jet's
momentum. An additional requirement of $0.9 < \mjets/\HT < 4$ is imposed,
consistent with Ref.~\cite{HION-2022-15}, to reduce contamination from jets not originating from the primary hard-scattering process.
\begin{figure}[!t]
\centerline{
\includegraphics[width=0.68\textwidth]{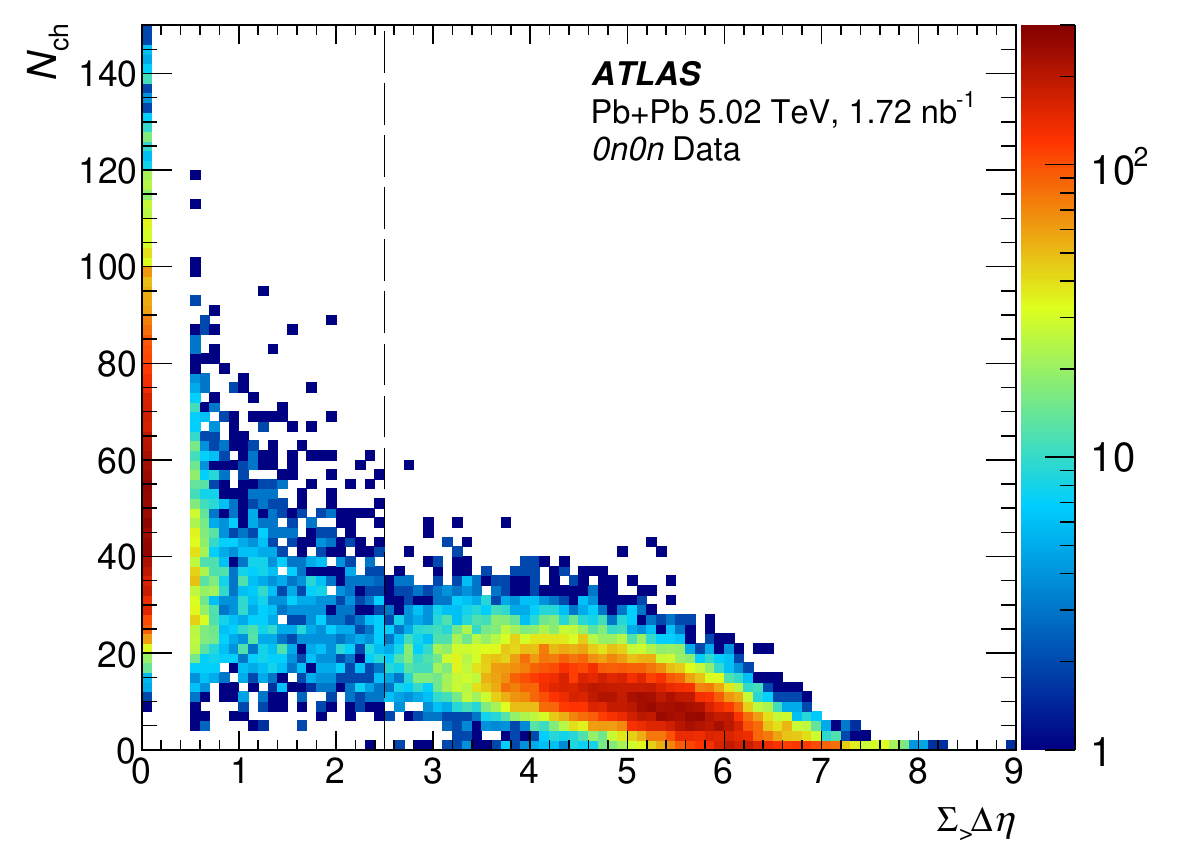}
}
\caption{
The correlation between the number of charged tracks (\Nch) and the maximum
single-sided sum of rapidity gaps (\sumGapMax) for events with at least two jets
passing all other trigger and fiducial requirements. The sample of events at
small \sumGapMax\ with a wide distribution in \Nch\ corresponds to the background
from hadronic \PbPb\ collisions, which is removed by the selection
$\sumGapMax > 2.5$.
}
\label{fig:gap_nch}
\end{figure}

The sum of rapidity gaps ($\sumGap$) is computed using the methodology
of Ref.~\cite{HION-2022-15} from a combination
of charged particle tracks and topo-clusters, both required to satisfy $\pT > 0.2$~\GeV.
Specifically, \sumGap is computed separately for
each side of the detector  as the sum of all rapidity gaps
greater than $0.5$ units between the edge of the detector and the
nearest jet. To exclude hadronic \PbPb\ collisions, i.e. typical collisions with nuclear overlap, the
larger of the sums between the two sides ($\sumGapMax$) is required to satisfy $\sumGapMax > 2.5$.
For a \znzn\ topology, this selection achieves a better rejection of hadronic
collisions than in \znxn\ measurements~\cite{HION-2022-15}
because hadronic \PbPb collisions rarely occur without forward neutron emission.
The impact of the \sumGapMax selection is shown in Figure~\ref{fig:gap_nch}, which
demonstrates the correlation between $\sumGapMax$ and the multiplicity of
charged tracks, \Nch, defined as the number of
reconstructed tracks satisfying $|\eta^\text{tr}| < 2.5$
and $\pT^\text{tr} > 0.2~\GeV$. The higher-multiplicity events at low \sumGapMax
and peaked at $\sumGapMax = 0$ correspond to the
hadronic \PbPb\ collision background removed by the selection at
$\sumGapMax > 2.5$.

To select \znzn\ collisions, a veto is applied on events having a
reconstructed  energy in either ZDC greater than
40\% of the beam energy per nucleon, i.e. more than 1~TeV. This
requirement augments and tightens the requirement applied to the
ZDCs in the trigger. All of the requirements used to select
\znzn\ UPC jet events are summarized in Table~\ref{tbl:select}.

\begin{table}[tb]
\centering
\caption{Jet and event selections applied in the measurement \znzn\ UPC jet events.
The quantities $E_\mathrm{A}$ and $E_\mathrm{C}$ are the ZDC energies on the A or C sides of the
ATLAS detector, respectively.
}
\begin{tabular}{l|l}\hline
Jet & Event \\ \hline
$\pTjet>15$~\GeV & ZDC \znzn: $E_\mathrm{A} < 1~\TeV$ and $E_\mathrm{C} < 1~\text{TeV}$ \\
$\etajet < 4.4$  &  $N_{\text{jet}} \geq 2$\\
& $0.9\HT < \mjets < 4\HT$\\
& $\sumGapMax > 2.5$\\
\hline
\end{tabular}
\label{tbl:select}
\end{table}

\subsection{Jet system kinematic variables}

Due to the dominance of coherent pomeron emission from the lead nucleus, the pomeron and photon energies in
\gamPomJets\ events are similar, yielding an approximately
symmetric $\yjets$ distribution, despite the asymmetric collision system
and gap topology. Thus, the cross-sections of \gamPomJets\ events are measured
as a function of the symmetric variable, $|\yjets|$. Non-diffractive \gamA collisions, however, are intrinsically
asymmetric, as the photon energies are usually much smaller than the longitudinal
momenta of the struck partons. Thus, to properly describe the physics of the
partonic scattering, it is useful to define jet kinematic variables that take advantage of
this asymmetry. In particular, for \znzn\ collisions, the energy asymmetry implies
that for sufficiently large $|\yjets|$, the nucleus-going side of the collision
may be identified with the direction of \yjets.
This assumption is valid for more than $99.9\%$ of the studied events with $|\yjets| > 1$
in the \gamAjets\ MC sample. Two momentum
fractions are defined as a function of $|\yjets|$ and \mjets:
\begin{align}
\zneg \equiv\frac{ \mjets }{\sqn} e^{-|\yjets|},~ \xpos \equiv\frac{ \mjets}{\sqn} e^{+|\yjets|}\, .
\end{align}
Neglecting additional radiation and contributions beyond leading-order, these
correspond to fractions of the beam momentum carried by the
hard-scattering participants from the photon-emitting nucleus (\zneg) and struck nucleus (\xpos). These variables are
analogous to \zg\ and \xA\ defined in Ref.~\cite{HION-2022-15} for collisions that
exhibit a \znxn\ topology. However, they differ from \zg\ and \xA\ since without
ZDC information to determine the photon-going direction of the
collision, the direction must be
inferred from the sign of $\yjets$.
\begin{figure}[!t]
\centerline{
\includegraphics[width=0.9\textwidth]{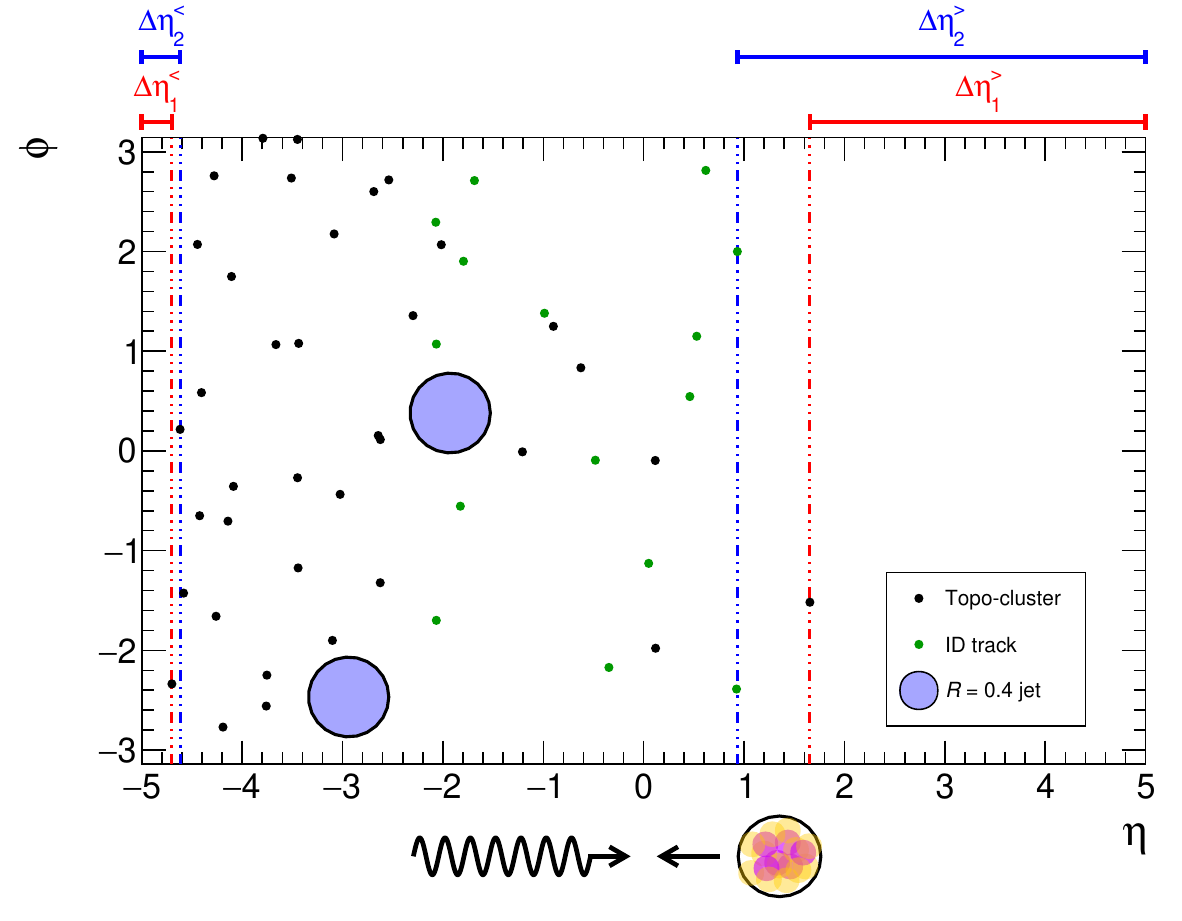}
}
\caption{
A diagram illustrating how \edgeGapTwoMin\ is calculated for a typical \gamAjets\
event. Tracks and topo-clusters are indicated by green and black points, respectively,
while jets are shown as blue circles. The $\Delta\eta$ variables computed on both sides
of the detector are shown, where $\Delta\eta_{1}$ is the gap from the detector edge
to the nearest track or topo-cluster, while \edgeGapTwo\ is the gap from the detector
edge to the second-closest track or topo-cluster.
}
\label{fig:GapCartoon}
\end{figure}

\subsection{Background subtraction}
The relative contributions of \gamA, \gamPom, and \gamgam\ processes
to the total \znzn\ cross-section are assessed via a template-fitting
procedure. To perform these fits, the single-sided
\edgeGapTwo\ variable is defined as the pseudorapidity interval
between the edge of the ATLAS detector ($\eta = \pm 4.9$) and the
second-closest track or topo-cluster on a given side of the
detector. Because tracks and topo-clusters resulting from particle
production rarely occur in isolation while clusters arising from
calorimetric noise often do -- see Figure~\ref{fig:GapCartoon} for an
illustration -- computing the rapidity gap relative to
the second-closest track or topo-cluster, rather than the closest,
reduces the sensitivity of the gap variable to noise, without losing
sensitivity to the physics.
After calculating \edgeGapTwo\ for both sides of the detector,
template fits are performed as a function of the minimum
\edgeGapTwo\ between the two detector sides, \edgeGapTwoMin.
These template fits are binned in two different sets of jet
kinematic variables: $(|\yjets|, \HT)$ for \gamPom\ measurements and
$(\zneg, \xpos)$ for measurements of the EMD rate in \gamA\ processes.

The gap distributions measured in
data are modeled as the sum of three components, corresponding
to the three processes that contribute to \znzn\ UPC jet production:
\gamAjets, \gamPomJets, and \gamGamJets.
The \gamAjets\ template is taken directly from \znxn\ data in the
corresponding kinematic interval and is treated as pure \gamAjets\ signal.
However, since the residual hadronic collision background is slightly larger in the
\znxn\ template and concentrated entirely at the lowest \edgeGapTwoMin\
values, the first bin of the data distribution is excluded from the fit. For the
\gamPomJets\ and \gamGamJets processes, no pure data-driven templates are available, so
the templates are obtained from the corresponding \pythia\ MC samples.
\begin{figure}[!p]
\centerline{
\includegraphics[width=0.99\textwidth]{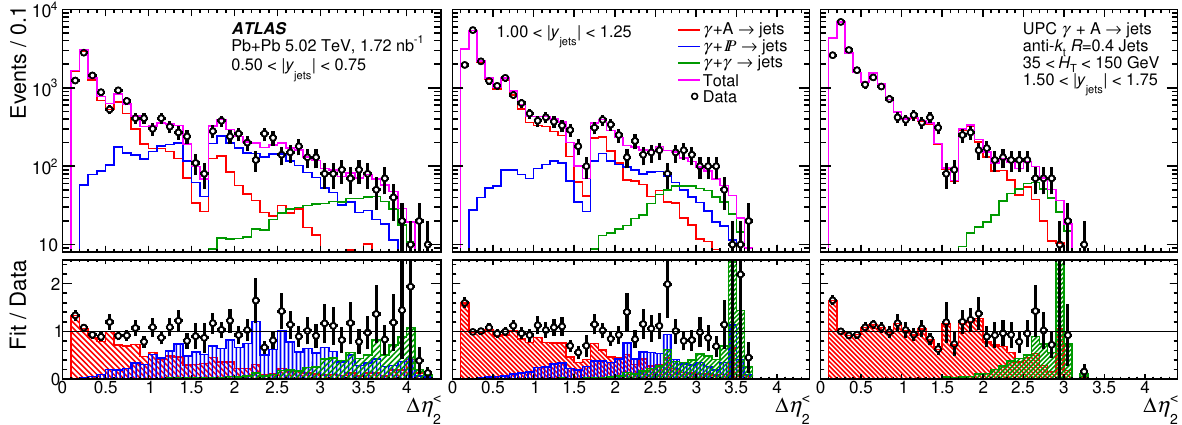}
}
\caption{
Example template fits of the \edgeGapTwoMin distribution for \znzn\ UPC jet events
passing all other selections in three
jet kinematic intervals: $0.5 < |\yjets| < 0.75$ (left), $1.0 < |\yjets| < 1.25$ (center),
and $1.5 < |\yjets| < 1.75$ (right). The lower
panel shows the ratio of different process templates to data, and
the markers show the ratio of the full template fit to the data. At large $|\yjets|$,
the best fit determines no \gamPom\ contribution and a small but non-negligible
\gamgam\ contribution.
}
\label{fig:tempfitbins}
\end{figure}

\begin{figure}[p]
\centerline{
\includegraphics[width=0.75\textwidth]{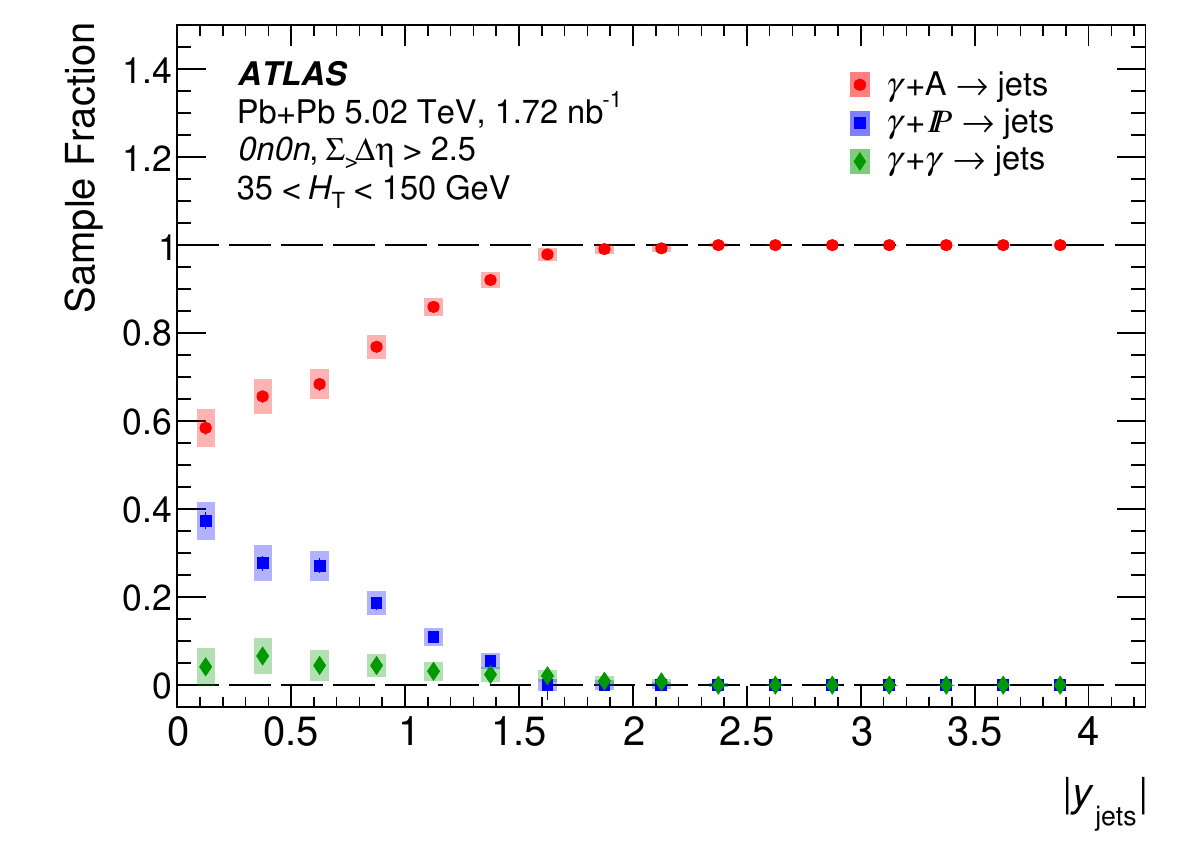}
}
\caption{
The fraction of jet events resulting from \gamA, \gamPom, and \gamgam processes,
extracted from template fits of the \edgeGapTwoMin distributions. Error bars
represent statistical uncertainties, and the shaded bands represent systematic
uncertainties. These fractions are reported for template fits integrated over
the full fiducial region in \HT\ in order to improve their statistical precision.
}
\label{fig:sigfracs_all}
\end{figure}

Figure~\ref{fig:tempfitbins} demonstrates the performance
of the \edgeGapTwoMin template-fitting procedure; each panel corresponds to a different
$|\yjets|$ bin, inclusive in \HT.
Several features are visible in the template fits, including a dip around
$\edgeGapTwoMin = 1.7$, corresponding to the transition between the ATLAS
forward calorimeter and hadronic endcap regions. The impacts of any imperfections
in the template-fitting procedure on the derived signal fractions are
assessed when determining systematic uncertainties, as discussed in
Section~\ref{sec:systematics}. The first bin of the fit is excluded
from the chi-squared calculation due to its different sensitivity to the cut-off at
the edge of the ATLAS detector. Figure~\ref{fig:sigfracs_all} shows the
signal fractions obtained from the template fits as a function of $|\yjets|$.
The results reflect the kinematics of the different processes:
\gamPomJets\ and \gamGamJets\ are more balanced in energy and thus
peaked near $\yjets = 0$, while \gamAjets\ is more imbalanced in
energy, causing it to dominate the forward region. These results also demonstrate
the reason for an upper bound at $|\yjets| = 1.5$ for the \gamPomJets\
measurement, namely that the backgrounds grow proportionally too large to
effectively extract the signal.

\begin{figure}[!tb]
\centerline{
\includegraphics[width=0.7\textwidth]{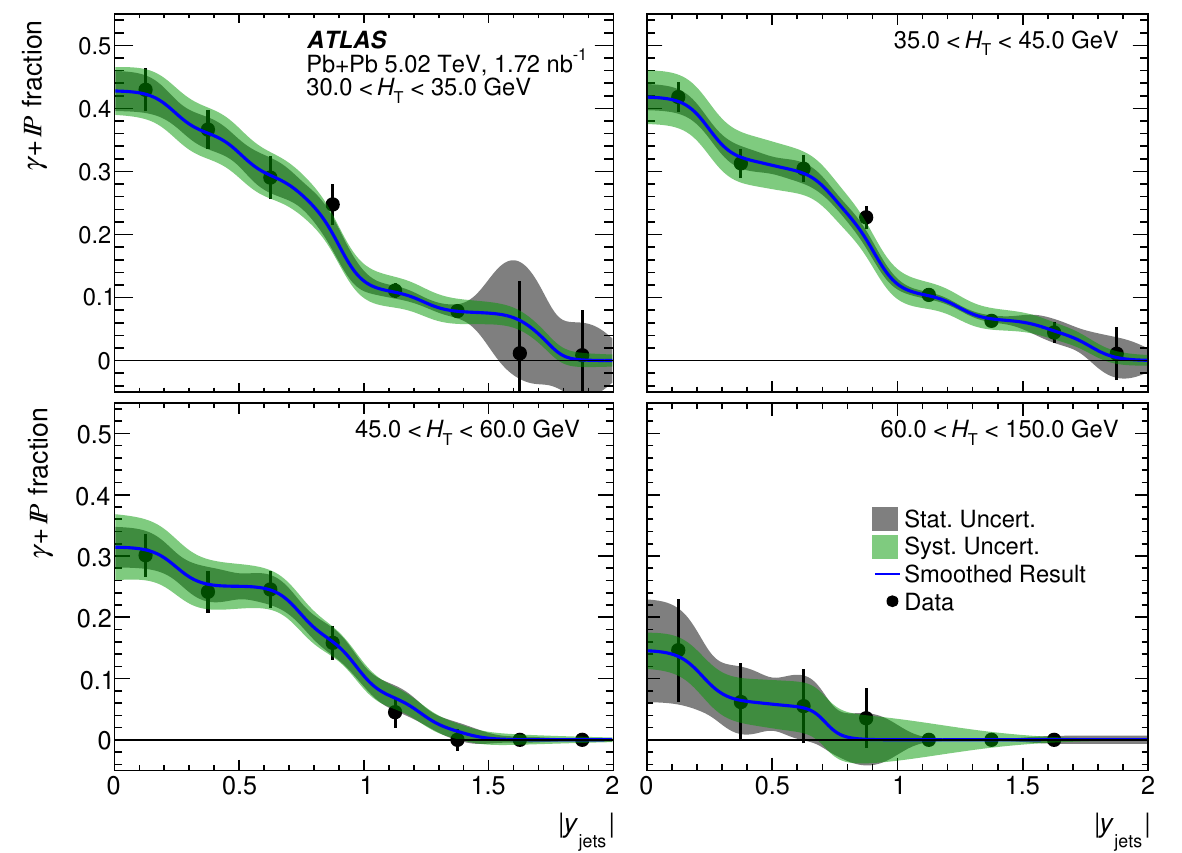}
}
\caption{
Fractional contribution of \gamPomJets\ processes to the total \znzn\ UPC jets cross-section
as a function of $|\yjets|$ in four different bins of \HT. While
the \gamPomJets\ cross-sections are not reported for $\HT < 35$~\GeV,
signal fractions are derived in this region for use in the
unfolding. The binned results from template fits are shown as
points, and the smoothed results (blue line) are produced using a Gaussian
kernel smoother in two dimensions. The statistical and systematic
uncertainties are similarly smoothed and shown as the black and
green shaded bands, respectively.
}
\label{fig:sigfracs_diff}
\end{figure}

After performing the template fits in bins of $|\yjets|$ and \HT\ for
the \gamPomJets\ measurement, the resulting signal fractions are smoothed using a
two-dimensional Gaussian kernel procedure. The  \gamPomJets\ signal
fractions obtained from the fits and the smoothed results
are shown as a function of $|\yjets|$ in bins of
\HT\ in Figure~\ref{fig:sigfracs_diff}. The smoothing allows the signal
fractions to be translated from the binning used for the template fits to
the bins used to compute the \gamPomJets\ cross-sections. These signal fractions generally
decrease with increasing $|\yjets|$ and \HT. The trend in $|\yjets|$
is explained by the different energy balance in \gamAjets\ events that dominate
at forward rapidity. The trend with \HT\ indicates that the \gamAjets\ \HT\
distribution falls less steeply than that of the \gamPomJets\ case.

\begin{figure}[!tb]
\centerline{
\includegraphics[width=0.6\textwidth]{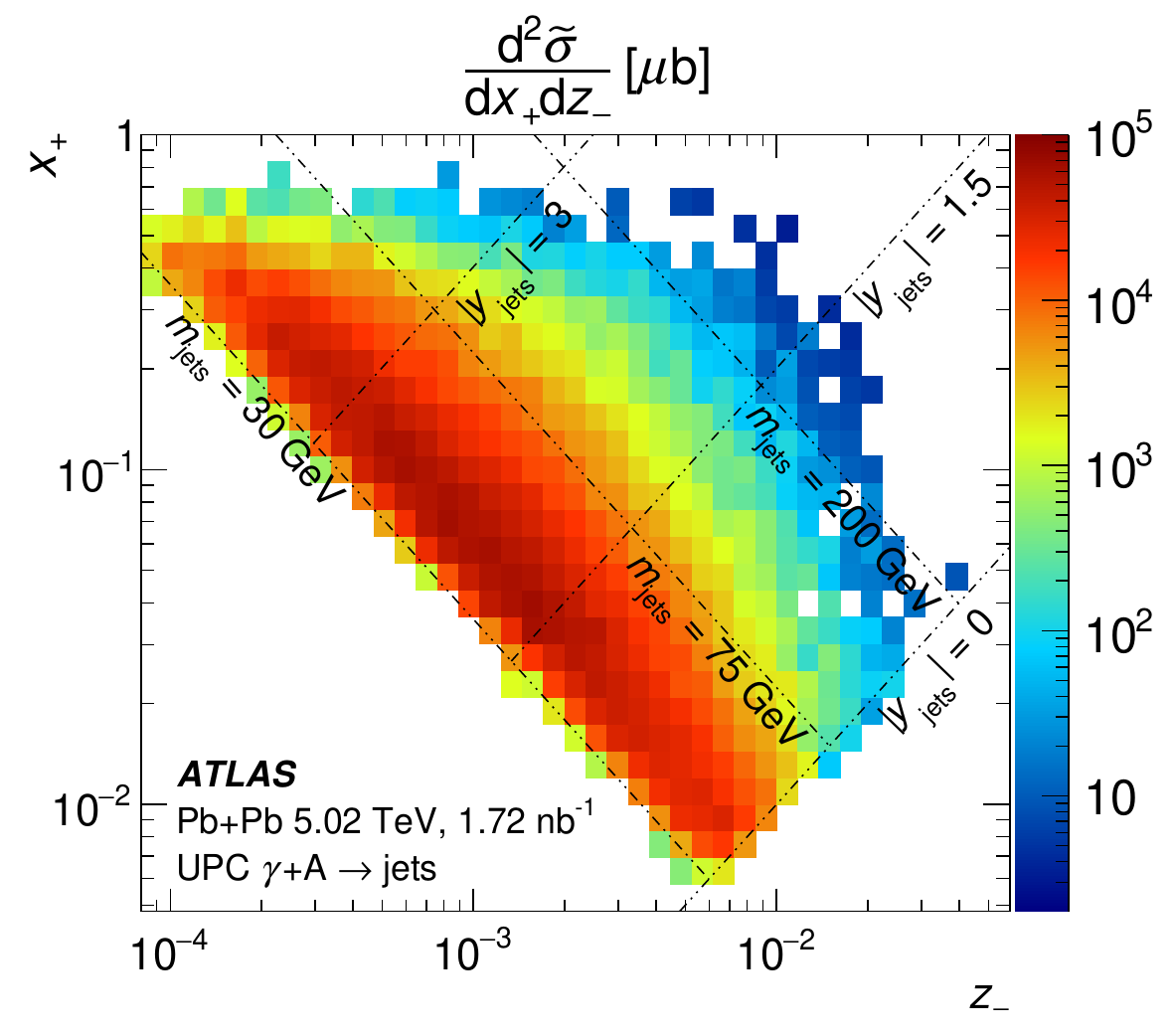}
}
\caption{
Fiducial cross-sections without unfolding corrections for
non-diffractive \gamAjets\ in the \znzn\ event class as a function of \xpos\ and \zneg.
The dashed lines indicate the correspondence between \xpos\ and \zneg\ for certain
fixed values of $|\yjets|$ or \mjets.
}
\label{fig:ResultsGamA}
\end{figure}

Figure~\ref{fig:ResultsGamA} shows the luminosity-normalized yields -- i.e.
cross-sections without unfolding corrections -- of \gamAjets\
with a \znzn\ topology as a function of \xpos\ and \zneg.
The dashed lines indicate
how different jet kinematics map onto the \xpos--\zneg space. The dashed line at
$\yjets=0$ in particular corresponds to $\xpos = \zneg$ and indicates the
constraint imposed in the definition of the variables $\xpos \geq \zneg$:
the photon-going direction is aligned with the direction
of \yjets. A more complete assessment of the unfolded results
and their ratio to the \znxn\ cross-section is provided in Ref.~\cite{OtherPaperPRL}.

\subsection{Data corrections and unfolding}
\begin{figure}[!b]
\centerline{
\subfloat[]{\includegraphics[width=0.49\textwidth]{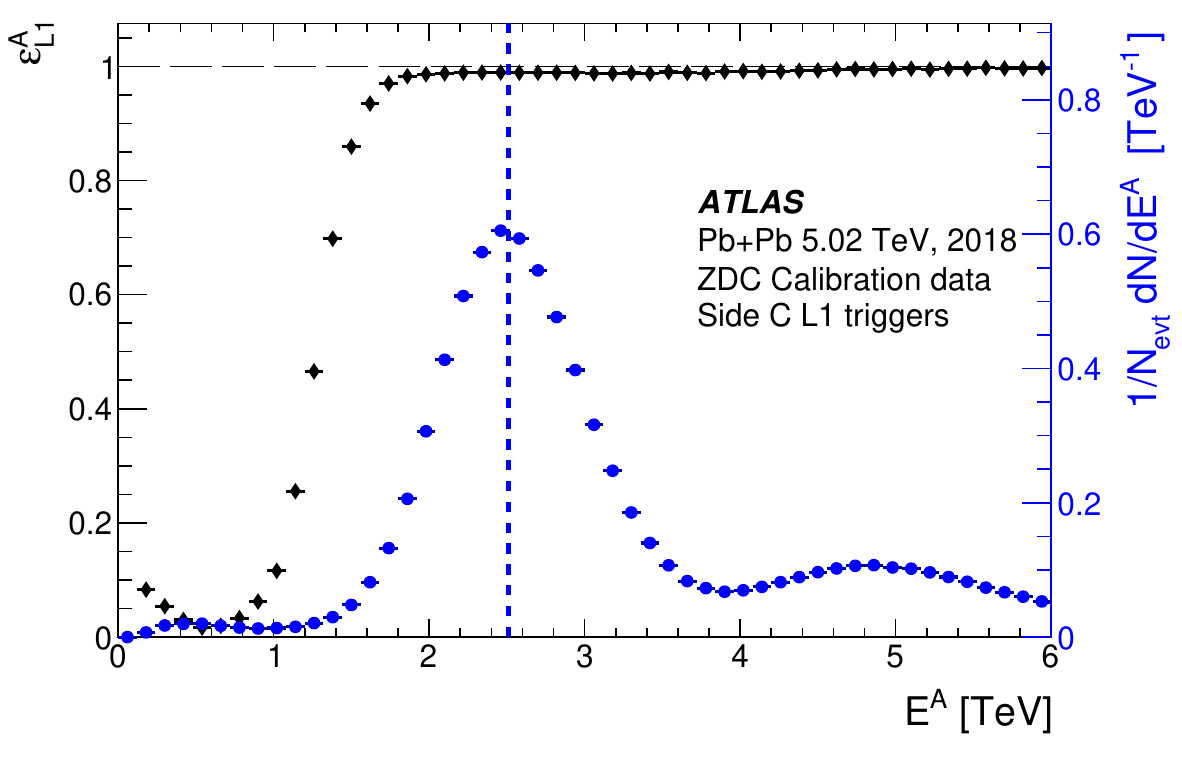}}
\subfloat[]{\includegraphics[width=0.49\textwidth]{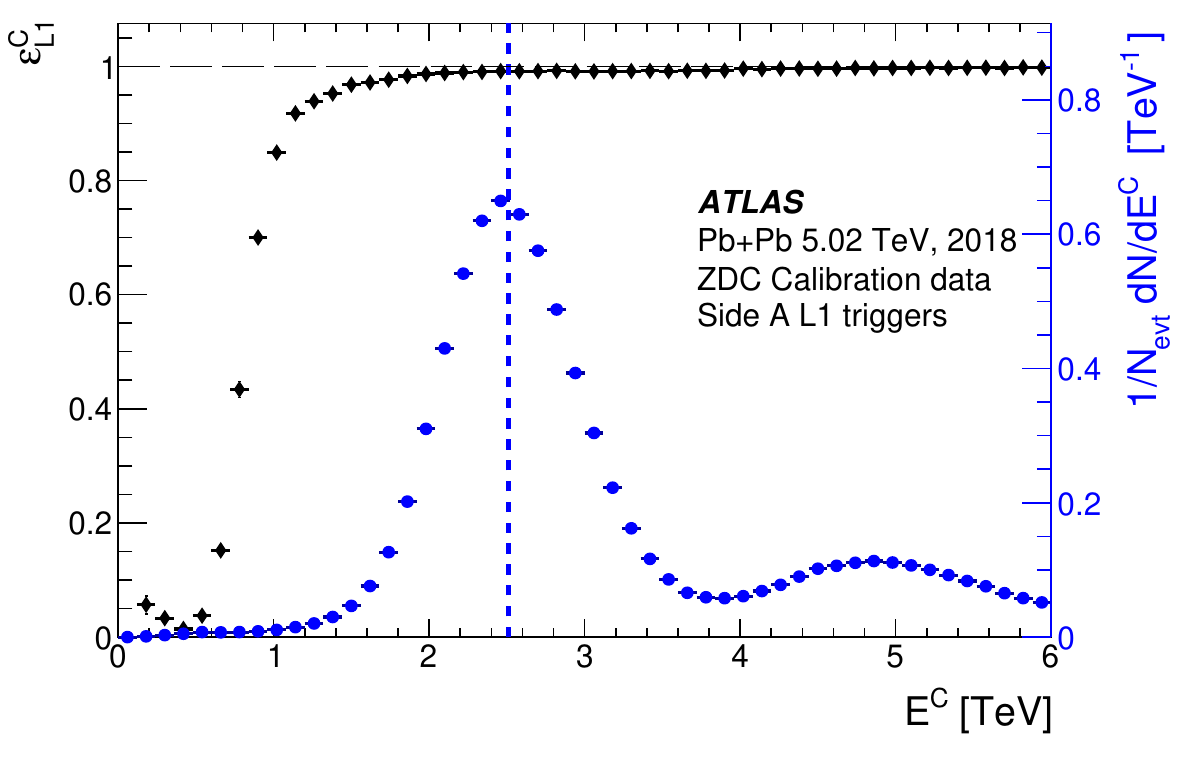}}
}
\caption{
The ZDC trigger efficiency (black) compared to the re-scaled ZDC
energy distribution (blue) for (a) the A-side and (b) the C-side of the
ATLAS detector. Results are derived from a dedicated high-statistics
sample of ZDC calibration data from the same data-taking period as the
other results in this paper. The energy distributions
are normalized by the total number of events plotted, and a dashed line
marks the beam energy per nucleon, $2.51$~\TeV.
The slight increase in the trigger efficiency
for small but non-zero energies results from residual activation of
the ZDC absorber that can produce pulses that fire the trigger but are
excluded from the offline energy measurement with the result that the
calibrated energy is small or zero. These events have negligible
impact on the \znzn measurement but appear in the efficiency due to
the extremely low rate of true low-energy events in the ZDC.
}
\label{fig:ZdcTrigEff}
\end{figure}

Event-by-event corrections are applied to the data to account for imperfect
efficiencies of the trigger and rapidity gap selections. Possible
trigger inefficiency arises from three sources:
the L1 ZDC selections, L1 \sumET\ thresholds, and HLT jet requirements.
Figure~\ref{fig:ZdcTrigEff} shows the ZDC trigger efficiency
as a function of ZDC energy for each side of the ATLAS detector (labeled sides A and C),
overlaid with the energy distribution
of neutrons. Here, sides A and C denote the $+z$ and $-z$ sides of
the ATLAS detector, in the coordinate system defined
in Section~\ref{sec:detector}. The ZDC trigger efficiency on
either side of the detector is studied using triggers which
select events in the ZDC on the opposite
side, derived from a high-statistics
sample of ZDC calibration data taken during the same period as
the data in this measurement. The offline selection to differentiate
$0n$ and $Xn$ topologies is 1~\TeV, corresponding to $40\%$ of the beam energy per
nucleon. These results indicate that for events with exactly one neutron, the trigger is
highly efficient ($\sim95\%$) and for all other events, it is fully efficient, yielding
an overall efficiency of greater than $99.9\%$ for selecting \znzn\ and \znxn\ events.
The efficiency of the L1 \sumET\ and jet requirements
rise rapidly at low \HT, such that both are at
least $98\%$ efficient in the fiducial region, $\HT > 35$~\GeV, for which a correction is applied. The
efficiency of the $\sumGapMax > 2.5$ requirement is
assessed using \pythia\ MC samples for non-diffractive
and diffractive events, and the data are corrected for any inefficiency. The
largest such correction in the fiducial region is less than $5\%$. An additional normalization
correction accounts for independent \PbPb\ EMD scatterings that
occur in the same crossing as an event of interest. Corrections are derived using
the methodology from Refs.~\cite{HION-2016-02,HION-2021-16}, and the resulting
correction to the normalization, averaged over all signal events, is $13(7)\%$ for the \znzn(\znxn) topology.

The fully corrected differential yields for the \gamAjets\ and
\gamPomJets samples are then obtained. They
are unfolded with a Bayesian unfolding procedure~\cite{DAgostini:1994fjx} implemented in
\verb+RooUnfold+~\cite{Adye:2011gm}, to correct for residual detector effects.
The unfolding of the \gamPomJets\ yields is performed in two dimensions as a function of $|\yjets|$ and
\HT, with the fiducial region consisting of six linear bins in $|\yjets|$ from $0$ to $1.5$
and four logarithmic bins from $35$ to $150$~GeV in \HT. For the reported cross-sections,
the last two \HT\ bins are merged to improve the statistical precision.
Unfolding corrections to the
\gamPom\ cross-sections vary between $30\%$ and $5\%$, decreasing at larger $|\yjets|$.
The \gamAjets\ yields used to compute the EMD fraction are unfolded three-dimensionally
in \HT, \xpos, and \zneg, using bins identical to those from the previous measurement of this
quantity described in Ref.~\cite{HION-2022-15}. These results are integrated over the same
wide regions in \xpos ($0.015 < \xpos < 0.2$) and \HT ($35 < \HT < 150$~\GeV) to compute the ratios.
Statistical uncertainties on the unfolded results are computed using $1000$ stochastic
variations of the data and response matrices.


%

%
\section{Electromagnetic dissociation}
\label{sec:emd}

In \gamAjets\ scattering processes with both nuclei nominally intact, the event's ZDC
topology may be modified by Coulomb excitation and subsequent decays of either nucleus, which
may produce additional neutrons. The fraction of events with such an EMD process where
the photon-emitting nucleus dissociates, $f^{\gamma}_\text{EMD}$, is
of particular interest. It serves as both a correction applied to the data and a method of
constraining the typical impact parameter relative to the center of the struck nucleus
($b_\mathrm{A}$) of \znzn\ \gamAjets\ events, due to its direct correlation with $b_\text{AA}$
(see Refs.~\cite{HION-2023-13,CMS-HIN-19-014,CMS-HIN-22-002,Guzey:2013jaa}).
To measure $f^{\gamma}_\text{EMD}$, one exploits the fact that events with
neutrons in the direction opposite to that indicated by \yjets\ are
dominated by peripheral \gamA\ processes in which the photon-emitting nucleus also emits
EMD neutrons. Since the direction of forward neutron emission in such events is
random, events with neutrons present in the same direction as that of the photon-emitting nucleus
reverse the ZDC topology of typical \znxn\ \gamAjets\ collisions, making them
distinguishable from the much larger ``true'' \znxn\ contribution.
These ``reverse'' events (\xnzn\ events) are selected
by requiring the sign of \yjets\ to be opposite to
the direction of the detected forward neutrons. To further reduce contamination
from ``true'' \znxn\ \gamAjets\ events, an additional
rapidity gap requirement is imposed of $\edgeGapTwoGam > 2.5, \edgeGapTwoA < 3$, where
\edgeGapTwoGam\ and \edgeGapTwoA\ align with the $0n$ and $Xn$ sides, respectively.
A sample of \znzn\ events without EMD is selected using the same
\edgeGapTwoGam\ and \edgeGapTwoA\ selections, and the fraction of peripheral
\gamAjets\ events accompanied by EMD is computed as:

\begin{equation}
f^{\gamma}_{\text{EMD}} \equiv
\frac{\left.\text{d}\sigma/\text{d}\zneg\right|_{Xn0n}^\text{rev.}}{\left.\text{d}\sigma/\text{d}\zneg\right|_{Xn0n}^\text{rev.}
+ \left.\text{d}\sigma/\text{d}\zneg\right|_{0n0n}},
\label{eq:fnoBU}
\end{equation}

where $\text{d}\sigma/\text{d}\zneg|_{\znzn}$ is the cross-section measured in the
\znzn\ sample as a function of \zneg\ and $\text{d}\sigma/\text{d}\zneg|_{\xnzn}^\text{rev.}$
is the cross-section measured in the \xnzn\ sample as a function of \zneg. Crucially,
since $\text{d}\sigma/\text{d}\zneg|_{Xn0n}^\text{rev.}$ is measured only for events with
neutrons on the side of the photon-emitting nucleus, it is sensitive
only to the half of EMD events where the photon-emitting
nucleus dissociates. This restriction makes the definition analogous (but not identical)
to $(1 - f_\text{no BU})$, where $f_\mathrm{no BU}$ is the no-breakup fraction measured
in Ref. \cite{HION-2022-15}. This quantity provides a measurement of the EMD rate of the
photon-emitting nucleus for events with a \znxn\ topology.

While the fraction of \gamAjets\ events where the photon-emitting nucleus dissociates
may be measured in both \znzn\ and \znxn\ collisions, the measured quantities are not directly comparable.
In the \znxn\ case, additional EMD of the photon-emitting nucleus transitions the event to an \xnxn\
topology, regardless of whether the struck nucleus also undergoes EMD. In the \znzn\ case, however,
the transition to \znxn\ occurs only when the photon-emitting nucleus \textit{alone} dissociates; events
where both nuclei dissociate (mutual EMD) cannot be measured, so they are not present in the numerator or
denominator of Equation~\ref{eq:fnoBU}. Thus, the \znxn\ measurement includes mutual EMD
while the \znzn\ measurement excludes it. While measuring correlated EMD of \znzn\ events
is impossible, a correction and associated uncertainties were derived to account for these
effects. This correction is applied to the \fBU\ measurements and theory predictions
for \znxn\ collisions. For further details on this procedure, see Appendix~\ref{sec:CorrEMD}.


%

%
\section{Systematic uncertainties}
\label{sec:systematics}

Systematic uncertainties on the measured cross-section ratios are assessed to account
for effects arising from the
event selections, jet energy measurement, sensitivity to the unfolding prior, and
signal fraction determination.
The event selection uncertainties are assessed for the gap requirement
($\sumGapMax > 2.5$) using the same procedure as Ref.~\cite{HION-2022-15}, where
it is tightened by 0.5 units to $\sumGapMax > 3.0$ and both the cross-sections
and efficiency corrections are re-computed accordingly.
The jet response uncertainties are propagated from the variations derived in
Ref.~\cite{HION-2022-15}, following the standard ATLAS jet calibration procedure for
Run 2~\cite{JETM-2018-05}. To derive these uncertainties, the jet energy is varied when constructing the response
matrices to account for uncertainties in the jet energy scale (JES) and resolution (JER). The results
unfolded with the varied response are compared to the nominal results.
An additional uncertainty on the jet energy scale is assigned to
account for the transfer of the \gamA\ jet calibration to the \gamPom\ system.
The uncertainty due to residual sensitivity to the unfolding prior is determined by
re-weighting the prior distribution in the response matrices to match the
reconstructed data. The data are then unfolded again with this modified
response matrix and compared to the nominal result.

Imperfections in the template fitting procedure used to extract the signal fractions shown in
Figure~\ref{fig:tempfitbins} introduce some systematic uncertainty on the cross-section measurement.
First, to assess sensitivity to possible mismodeling in the tails of the distributions,
the fit range is varied by removing two additional bins at the low end in $\Delta\eta_2^{<}$ or five
bins at the upper end. Uncertainties
are also applied to account for imperfect modeling of the topo-cluster reconstruction
in simulation. An additional variation is applied where the proportion of direct and resolved photon
events in the $\gamma+\gamma$ template is varied to match the fractions observed in Ref.~\cite{HION-2022-15}.
An additional uncertainty is assigned to account for potential
mismodeling of the \gamPom\ template shape by re-weighting the $t$ distribution corresponding
to variations in the $B$ parameter from Equation~\ref{eqn:PomFlux}.
The $B$ parameter is expected to scale as the square of the ratio of the pomeron emitter radius,
as observed in the slopes of elastic diffractive $\rho^0$ photoproduction measured by H1~\cite{H1:2020lzc}
and STAR~\cite{STAR:2017enh} with a proton or gold target, respectively. Such a scaling by the radii
$R_\mathrm{Pb}/R_\mathrm{p} = 6.5$ yields a value of $B = 235$~$\GeV^{-2}$ that differs from
$B_\mathrm{nom} = 200$~GeV$^{-2}$, derived from fitting the nuclear form factor,
by $\Delta B = 35$~$\GeV^{-2}$. Since $B_\mathrm{nom}$ is found to better
describe the shape of the $|\yjets|$ distribution in data, it is used for the nominal results. The uncertainty
on modeling the pomeron flux in hard diffraction from nuclei is assessed by varying $B$ to $165(270)$~GeV$^{-2}$
for the upward (downward) direction.

The uncertainties on the signal fractions are
propagated to re-derive cross-sections for each process, which are
then unfolded and compared to the nominal distribution to estimate the systematic
uncertainties. While the uncertainty is typically smaller than the systematic
uncertainties arising from the jet response near mid-rapidity, it grows in
the forward region, becoming comparable to or larger than the JES uncertainty. The two largest contributors to
the signal fraction uncertainty are the lower truncation in the fit range and the
diffractive shape variation. The dominant source of
signal fraction uncertainty in this region is imperfect modeling of the shape of
the \gamPom\ \edgeGapTwoMin\ distribution, which impacts the relative contribution
of \gamA\ and \gamPom\ processes, particularly at large $|\yjets|$.

\begin{figure}[!tb]
\centerline{
\includegraphics[width=0.65\textwidth]{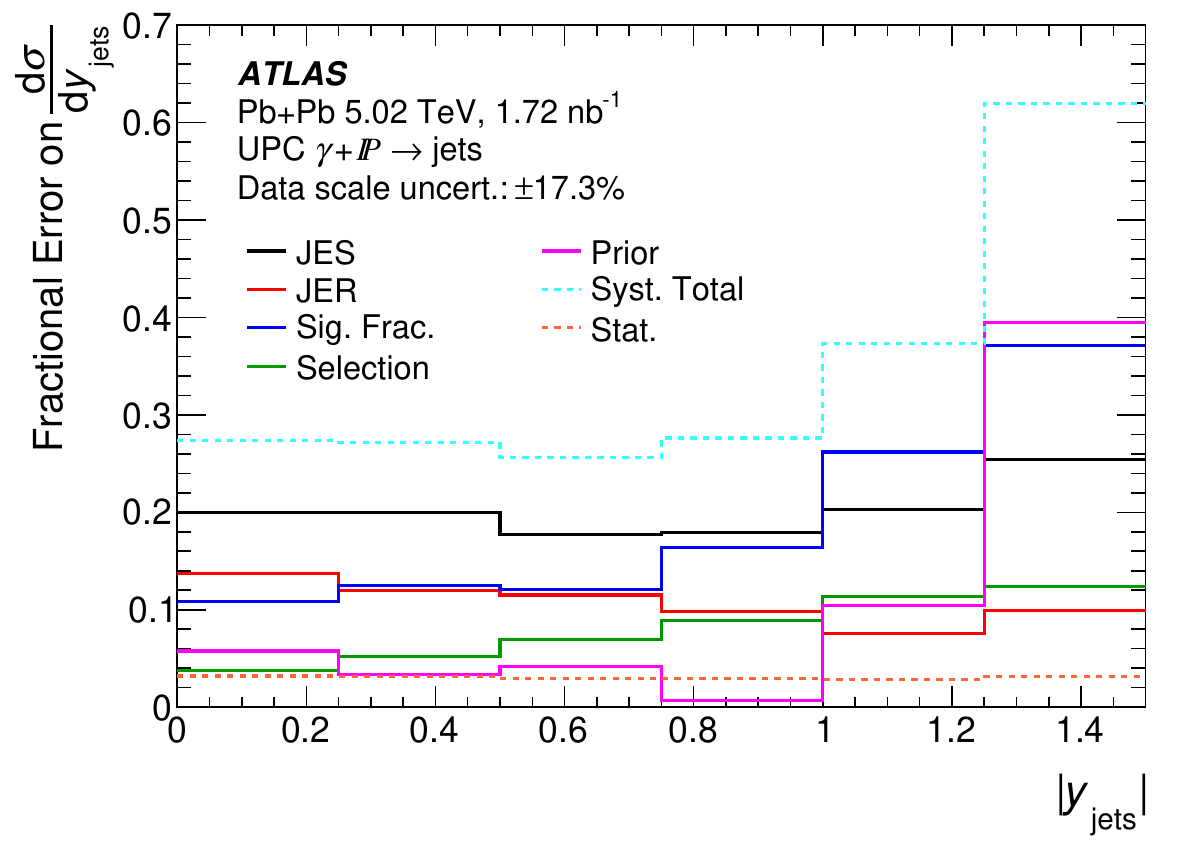}
}
\caption{
The fractional uncertainties on the \gamPomJets\ cross-sections as a function
of $|\yjets|$ from different
sources of systematic uncertainty which contribute in this measurement.
The jet energy scale systematic is dominant near mid-rapidity, but larger
uncertainties on the signal fraction overtake it in the forward region.
The fully correlated scale uncertainty is
$\pm 17.3\%$ is composed of $\pm14.5\%$ (JES) $\pm6.7\%$ (JER) $\pm6.2\%$ (Sig. Frac.)
$\pm 1.2\%$ (EMD pile-up) $\pm 2.0\%$ (Luminosity).
}
\label{fig:SystBreakdown}
\end{figure}

Figure~\ref{fig:SystBreakdown} shows how the different sources of systematic
uncertainty contribute over the range in $|\yjets|$. The dominant systematic
uncertainties arise from the jet energy scale measurement, where the uncertainty
on the jet calibration transfer from \gamAjets\ to \gamPomJets, assessed as
the full difference in MC-derived calibration factors between the samples, is the largest
component. In the forward region, the signal fraction uncertainty is largest,
while the systematic uncertainties on the jet energy scale remain substantial.
While Figure~\ref{fig:SystBreakdown} shows the uncertainties integrated in \HT,
Figure~\ref{fig:sigfracs_diff} indicates that the signal fraction uncertainty
grows for large \HT, because the contribution from \gamPomJets\ processes
becomes more challenging to constrain in kinematic regions with fewer events.
The jet response uncertainty components, aside from the calibration transfer,
and the diffractive shape component of the signal fraction uncertainty are
treated as fully correlated between bins. All other components
are treated as uncorrelated. Two additional components of fully
correlated uncertainty impact the overall normalization: the
luminosity measurement and the correction for EMD pile-up. The luminosity
uncertainty for the full 2018 Pb+Pb dataset is $2.0\%$, derived using methods
described in Ref. \cite{DAPR-2021-01}, using the LUCID-2 detector for the baseline luminosity
measurements \cite{LUCID2}. The uncertainty on the absolute cross-section arising due to the
correction for EMD pile-up is obtained by varying the \PbPb\ dissociative cross-section within
its uncertainties~\cite{PhysRevLett.109.252302}. This variation yields a
fractional uncertainty of $1.2\%$.

%
%
%
%
%
%
%
%
%
%
%
%
%
%
%
%
%
%
%
%
%
%
%
%
%
%
%
%
%
%
%


%

%
%

%
\section{Results}
\label{sec:results}

\begin{figure}[!tb]
\centerline{
\includegraphics[width=0.6\textwidth]{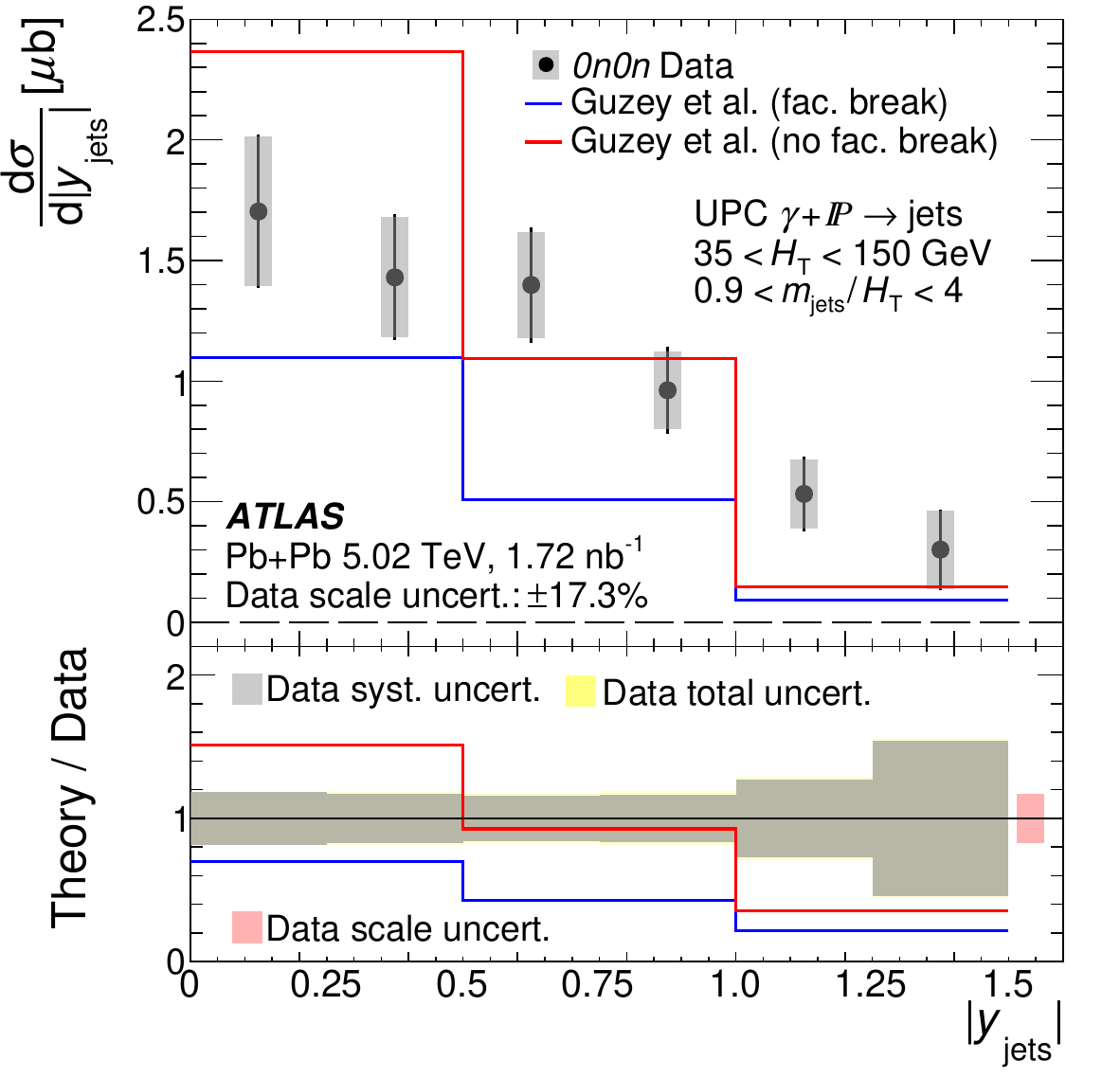}
}
\caption{
The \HT-integrated cross-sections for \gamPomJets\ production which are fully unfolded to
correct for detector response. The systematic uncertainties are shown as
shaded bands, and vertical lines represent the quadrature sum of the
statistical and systematic uncertainties. The fully correlated component
of the systematic uncertainty is subtracted from the bin-to-bin
contributions and denoted by the data scale uncertainty. NLO
theory predictions~\cite{Guzey:2016tek} are shown as blue
or red lines. The bottom panel shows the ratio of the different predictions to the data,
with total and systematic uncertainties shown as shaded yellow or gray bands,
respectively, where the yellow is nearly identical to the gray in most bins.
Ratios are computed by averaging the results of two bins in
order to take the ratio with respect to the less granular theoretical predictions.
}
\label{fig:ResultsDiff}
\end{figure}

\begin{figure}[!tb]
\centerline{
\includegraphics[width=0.99\textwidth]{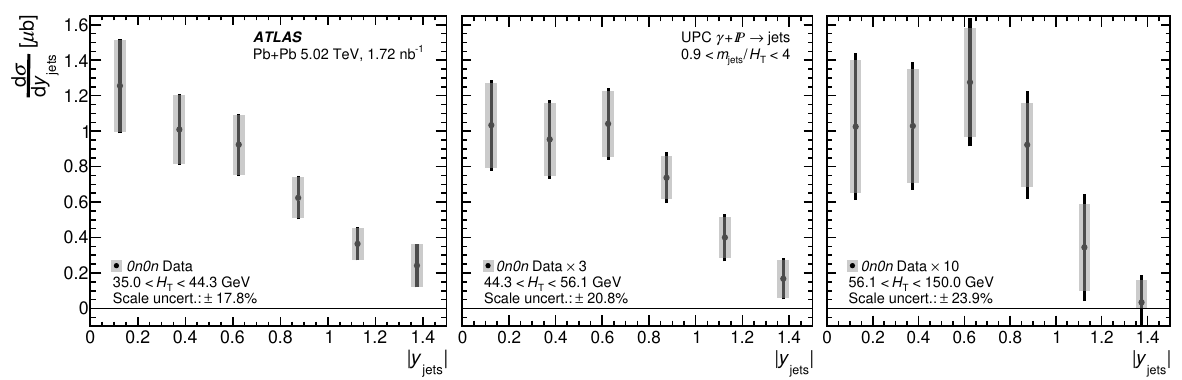}
}
\caption{
The \HT-differential cross-sections for \gamPomJets\ production which are fully unfolded to
correct for detector response. The systematic uncertainties are shown as
shaded bands, and the combined statistical and systematic uncertainties are shown as vertical lines.
The fully correlated component of the systematic uncertainty is subtracted from the bin-to-bin
contributions and denoted by the data scale uncertainty.
}
\label{fig:ResultsDiffHT}
\end{figure}

The fully unfolded \HT-integrated cross-sections for coherent \gamPomJets\
production  as a function of $|\yjets|$ are shown  in Figure~\ref{fig:ResultsDiff},
and the \HT-differential cross-sections are shown in Figure~\ref{fig:ResultsDiffHT}.
Both results are measured over the range $|\yjets| < 1.5$ and provide the first measurement
of diffractive jet production in coherent photonuclear interactions. The
$|\yjets|$-differential cross-sections are compared to next-to-leading-order
(NLO) theoretical predictions~\cite{Guzey:2016tek}, calculated in two different
scenarios for the strength of factorization-breaking~\cite{Klasen:2004qr}
effects. The shape of the data distribution differs from the predictions, indicating that
the modeling of the coherent pomeron flux may require modification before quantitative
conclusions about factorization-breaking can be drawn from these data. Improved
modeling of the pomeron flux may also provide sensitivity to other phenomena
influencing these cross-sections, such as the pomeron PDFs. The relative systematic uncertainty
increases for $|\yjets| > 1$, particularly at large \HT, due to difficulty
constraining the relative contributions of the \gamA\ and \gamPom\ contributions in intervals
with few \gamPom\ events.

\begin{figure}[!tb]
\centerline{
\includegraphics[width=0.6\textwidth]{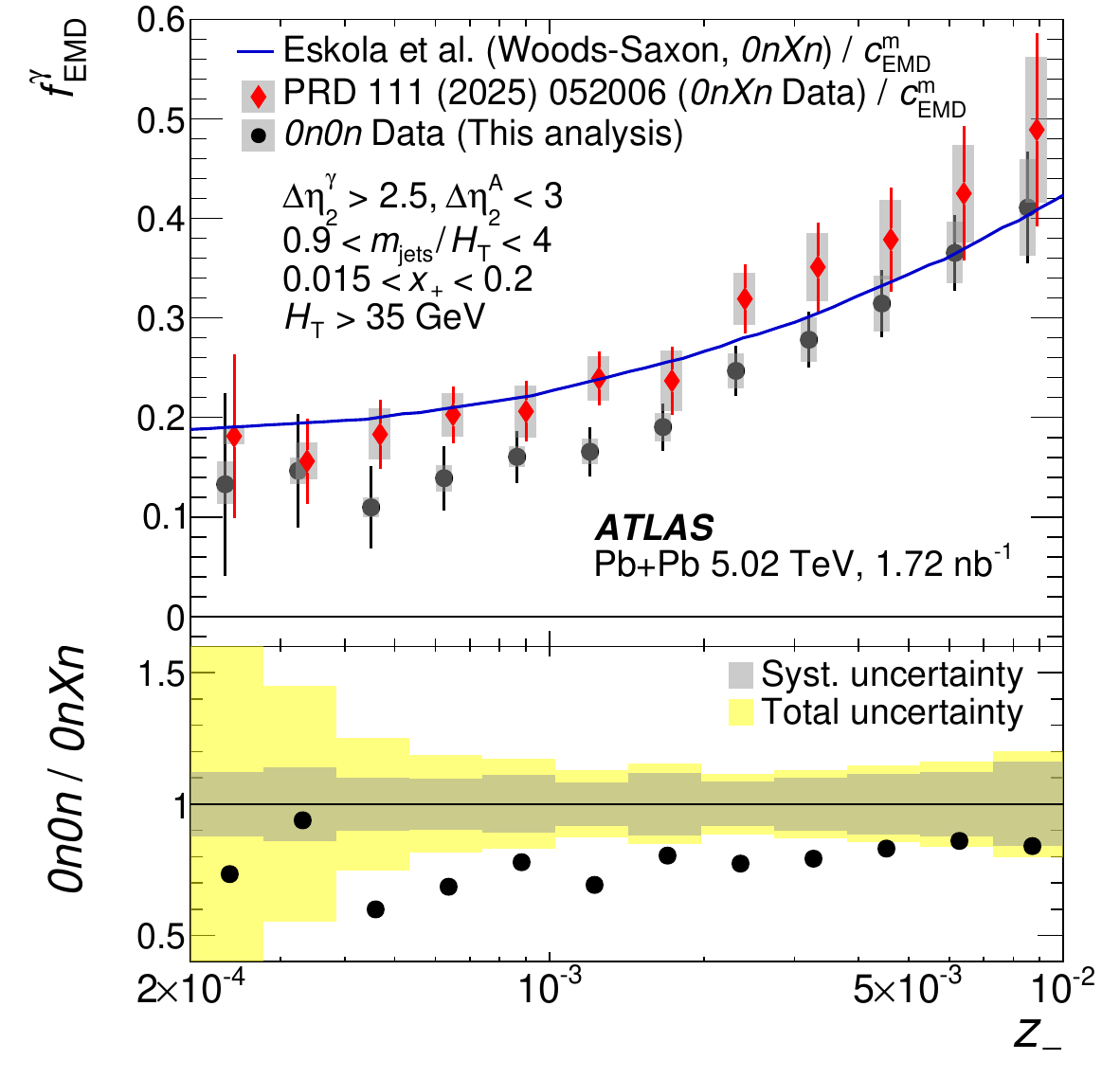}
}
\caption{
The fraction of \gamAjets\ events
in which the photon-emitting nucleus breaks up due to EMD as a function of \zneg.
The systematic uncertainties are shown as shaded gray bands, and the error bars
are combined statistical and systematic uncertainties. Results for the \znzn\ topology (black) are
compared to the \znxn\ measurement from Ref. \cite{HION-2022-15} (red).
Both sets of markers are shifted slightly to enhance legibility.
The bottom panel shows the ratio of the \znzn\ and \znxn\ measurements,
with total and systematic uncertainties shown as shaded yellow or gray bands, respectively.
Theoretical calculations for the \znxn\ sample are also shown from Ref. \cite{Eskola:2024fhf} as a blue line.
Both the \znxn\ data and theory calculation are corrected for correlated EMD effects.
}
\label{fig:ResultsEMD}
\end{figure}

Figure~\ref{fig:ResultsEMD} compares the fully unfolded $f^\gamma_\text{EMD}$
for \znzn\ collisions with the corresponding quantity
measured for \znxn\ collisions in Ref.~\cite{HION-2022-15}. For a proper comparison, the \znxn\
fractions are corrected for the impact of correlated EMD on the ratio (\cCorr), as
described in Section~\ref{sec:emd}. An additional theoretical uncertainty is
also applied to the \znxn\ data due to uncertainty on \cCorr.
These results indicate that the \znzn\ events exhibit a lower overall rate for EMD
than the \znxn\ events.
Since the EMD rate is correlated with the sampled
$b_\mathrm{AA}$~\cite{Klein:2016yzr,CMS-HIN-19-014},
this observation provides evidence for the hypothesis that \znzn\ collisions probe
larger typical $b_\mathrm{AA}$ than \znxn\ collisions.
Since the \znzn\ selection would not be expected to
modify the photon flux, this
observation also supports the interpretation that \znzn\ events
select peripheral \gamA\ collisions (larger $b_\mathrm{A}$). This result
is particularly important for the interpretation of results in Ref.~\cite{OtherPaperPRL},
where detailed comparisons of the \znzn\ and \znxn\ \gamAjets\ cross-sections
are used to probe the dependence of nPDF modifications on $b_\mathrm{A}$.

\FloatBarrier


%

%
%
%

%
\section{Conclusions}
\label{sec:conclusion}

In summary, ATLAS has studied the cross-section for photoproduction of
jets in ultra-peripheral 5.02~\TeV \PbPb collisions in which both
nuclei remain intact. The lack of nuclear
breakup is determined by the absence of beam-rapidity neutrons in
ZDCs. The \gamAjets, \gamPomJets, and \gamGamJets\ contributions to the \znzn\
cross-section are statistically separated using a template fit to
rapidity gap distributions. This procedure enables the first measurement of the
inclusive \gamPomJets\ cross-section in nuclear collisions. The \gamPomJets\
cross-section is presented as a function of the jet system rapidity
and compared to predictions from NLO perturbative QCD calculations for
different scenarios for the strength of factorization-breaking. The theoretical
comparisons capture the overall scale of the cross-section well, but
the data shows a somewhat broader $|\yjets|$ distribution than the predictions. The
evolution of the cross-section with \HT\ is measured, in addition to the \HT-integrated
$|\yjets|$-dependence, to provide additional information for improving theoretical models.

The rate of electromagnetic dissociation of the photon-emitting nucleus is measured for \znzn\ collisions
and compared to the corresponding rate in \znxn\ collisions, after correcting for the effects of correlated
mutual EMD. The \znzn\ events exhibit a systematically lower EMD rate than \znxn\ events, providing evidence for the
hypothesis that \znzn\ collisions probe larger typical nucleus-nucleus impact parameters. Since the \znzn\
selection does not modify the photon flux, this also supports the interpretation that \znzn\ collisions
preferentially strike nucleons near the edge of the target nucleus, providing sensitivity to the
impact parameter dependence of nuclear parton distributions.

These results demonstrate the potential of \znzn\ ultra-peripheral collisions as a tool for studying
both diffractive and peripheral non-diffractive \gamA\ processes. While these results provide a first
measurement of \gamPomJets\ processes in heavy ion collisions at the LHC, future measurements with
improved statistical and systematic precision from both LHC Run 3 and the HL-LHC may enable more detailed studies of
factorization-breaking in coherent nuclear diffraction. These phenomena will be explored even
further at the EIC, where it will be possible to exploit the ion-going neutron topology to study
diffractive and peripheral non-diffractive \gamA\ processes in even greater detail. These measurements
provide both an important step forward in understanding hard diffraction in nuclei and a critical test for a new
method of probing the dependence of nPDF modifications on impact parameter.

%
%
%
%


%

%
%
%
%
%
\section*{Acknowledgements}
%

%
%
%
%

%
%

%
%

We thank CERN for the very successful operation of the LHC and its injectors, as well as the support staff at
CERN and at our institutions worldwide without whom ATLAS could not be operated efficiently.

The crucial computing support from all WLCG partners is acknowledged gratefully, in particular from CERN, the ATLAS Tier-1 facilities at TRIUMF/SFU (Canada), NDGF (Denmark, Norway, Sweden), CC-IN2P3 (France), KIT/GridKA (Germany), INFN-CNAF (Italy), NL-T1 (Netherlands), PIC (Spain), RAL (UK) and BNL (USA), the Tier-2 facilities worldwide and large non-WLCG resource providers. Major contributors of computing resources are listed in Ref.~\cite{ATL-SOFT-PUB-2026-001}.

We gratefully acknowledge the support of ANPCyT, Argentina; YerPhI, Armenia; ARC, Australia; BMWFW and FWF, Austria; ANAS, Azerbaijan; CNPq and FAPESP, Brazil; NSERC, NRC and CFI, Canada; CERN; ANID, Chile; CAS, MOST and NSFC, China; Minciencias, Colombia; MEYS CR, Czech Republic; DNRF and DNSRC, Denmark; IN2P3-CNRS and CEA-DRF/IRFU, France; SRNSFG, Georgia; BMFTR, HGF and MPG, Germany; GSRI, Greece; RGC and Hong Kong SAR, China; ICHEP and Academy of Sciences and Humanities, Israel; INFN, Italy; MEXT and JSPS, Japan; CNRST, Morocco; NWO, Netherlands; RCN, Norway; MNiSW, Poland; FCT, Portugal; MNE/IFA, Romania; MSTDI, Serbia; MSSR, Slovakia; ARIS and MVZI, Slovenia; DSI/NRF, South Africa; MICIU/AEI, Spain; SRC and Wallenberg Foundation, Sweden; SERI, SNSF and Cantons of Bern and Geneva, Switzerland; NSTC, Taipei; TENMAK, T\"urkiye; STFC/UKRI, United Kingdom; DOE and NSF, United States of America.

Individual groups and members have received support from BCKDF, CANARIE, CRC and DRAC, Canada; CERN-CZ, FORTE and PRIMUS, Czech Republic; COST, ERC, ERDF, Horizon 2020 and Marie Sk{\l}odowska-Curie Actions, European Union; Investissements d'Avenir Labex, Investissements d'Avenir Idex and ANR, France; DFG and AvH Foundation, Germany; Herakleitos, Thales and Aristeia programmes co-financed by EU-ESF and the Greek NSRF, Greece; BSF-NSF and MINERVA, Israel; NCN and NAWA, Poland; La Caixa Banking Foundation, CERCA and AGAUR programs from Generalitat de Catalunya and PROMETEO and GenT Programmes Generalitat Valenciana, Spain; G\"{o}ran Gustafssons Stiftelse, Sweden; The Royal Society and Leverhulme Trust, United Kingdom; Eric and Wendy Schmidt Fund for Strategic Innovation, United States of America.

In addition, individual members wish to acknowledge support from Chile: Agencia Nacional de Investigaci\'on y Desarrollo (ANID FONDECYT reg. 1230987, FONDECYT 1230812, FONDECYT 1240864, Fondecyt 3240661, Fondecyt Regular 1240721); China: Chinese Ministry of Science and Technology (MOST-2023YFA1605700, MOST-2023YFA1609300), National Natural Science Foundation of China (NSFC 12275265, NSFC-W2543005); Czech Republic: Czech Science Foundation (GACR - 24-11373S), Ministry of Education Youth and Sports (ERC-CZ-LL2327, FORTE CZ.02.01.01/00/22\_008/0004632), PRIMUS Research Programme (PRIMUS/21/SCI/017); EU: H2020 European Research Council (ERC - 101002463); European Union: European Research Council (BARD No. 101116429, ERC - 948254, ERC 101089007), European Regional Development Fund (HE COFUND GA No.101081355, ERDF), Marie Sklodowska-Curie Actions (GAP-101168829); France: Agence Nationale de la Recherche (ANR-21-CE31-0013, ANR-22-EDIR-0002, ANR-24-CE31-0504-01); Germany: Deutsche Forschungsgemeinschaft (DFG - 469666862); China: Research Grants Council (GRF); Italy: Ministero dell'Università e della Ricerca (NextGenEU I53D23000820006 M4C2.1.1, SOE2024\_0000023); Japan: Japan Society for the Promotion of Science (JSPS KAKENHI  JP25H0063, JSPS KAKENHI JP22H01227, JSPS KAKENHI JP22H04944, JSPS KAKENHI JP22KK0227, JSPS KAKENHI JP24K23939, JSPS KAKENHI JP24KK0251, JSPS KAKENHI JP25H00650, JSPS KAKENHI JP25H01291, JSPS KAKENHI JP25K01011, JSPS KAKENHI JP25K01023); Poland: Polish National Science Centre (NCN 2021/42/E/ST2/00350, NCN OPUS 2023/51/B/ST2/02507, NCN OPUS nr 2022/47/B/ST2/03059, NCN UMO-2019/34/E/ST2/00393, UMO-2022/47/O/ST2/00148, UMO-2023/49/B/ST2/04085, UMO-2023/51/B/ST2/00920, UMO-2024/53/N/ST2/00869); Spain: Agència de Gestió d'Ajuts Universitaris i de Recerca. (AGAUR - 2023 BP 00141), Ministry of Science and Innovation (RYC2019-028510-I, RYC2020-030254-I, RYC2021-031273-I, RYC2022-038164-I), Ministerio de Ciencia, Innovación y Universidades/Agencia Estatal de Investigaci\'on (EU NextGenerationEU (PRTR-C17.I1), PID2022-142604OB-C22); Sweden: Carl Trygger Foundation (Carl Trygger Foundation CTS 22:2312), Swedish Research Council (Swedish Research Council 2023-04654, VR 2021-03651, VR 2022-03845, VR 2022-04683, VR 2023-03403, VR 2024-05451, VR 2025-05940), Knut and Alice Wallenberg Foundation (KAW 2023.0366); Switzerland: Swiss National Science Foundation (SNSF - PCEFP2\_194658); United Kingdom: The Binks Trust, Royal Society (NIF-R1-231091); United States of America: U.S. Department of Energy (ECA DE-AC02-76SF00515), John Templeton Foundation (John Templeton Foundation 63206), Neubauer Family Foundation.

%
%


%
\clearpage
\appendix
\part*{Appendix}
\addcontentsline{toc}{part}{Appendix}
\section{Correction for correlated electromagnetic dissociation}
\label{sec:CorrEMD}

As discussed in Section~\ref{sec:intro}, a comparison of EMD breakup
fractions ($f^\gamma_\mathrm{EMD}$) in \znzn\ and \znxn\ processes
may provide valuable insight on the (relative) $b_\mathrm{AA}$
distributions for the two processes. However, the EMD rate measured
in \znzn\ collisions is computed for events where \textit{only} the
photon-emitting nucleus dissociates, while for \znxn\ collisions, it
is computed for events where the photon-emitting nucleus
dissociates, \textit{irrespective of whether the other nucleus would
undergo electromagnetic dissociation}. While it is impossible to measure the
exact same quantity in \znzn\ and \znxn\ collisions, a correction for the
difference in the processes accounted for in $f^\gamma_\mathrm{EMD}$
may be derived. Two approaches are used to estimate the impact of mutual EMD:
a phenomenological approach deriving corrections from
$\gamma+\gamma\rightarrow\ell^++\ell^-$
measurements~\cite{HION-2016-02,HION-2021-16} and a theory-driven
calculation of mutual EMD in \gamAjets\ events.

In the first approach, the desired correction is estimated
by assuming that the rate
for additional EMD in $\gamma+\gamma\rightarrow\ell^++\ell^-$ processes
primarily reflects the typical $b_\text{AA}$ of the
collision.  Under this assumption, collisions with similar \fBU\
sample the same $\langle b_\text{AA}\rangle$ values and, thus, experience the same
correlated EMD effects, irrespective of the hard process. Then, it is
possible to derive a correction factor (\cCorr) to \fBU\ in \znxn\
collisions which allows it to be directly compared to the analogous
quantity in \znzn\ collisions. Such a correction factor is defined as:

\begin{equation}
\fBU = \frac{\fBUznxn}{\cCorr(\fBUznzn)}
\end{equation}

\begin{figure}[!t]
\centerline{
\includegraphics[width=0.7\textwidth]{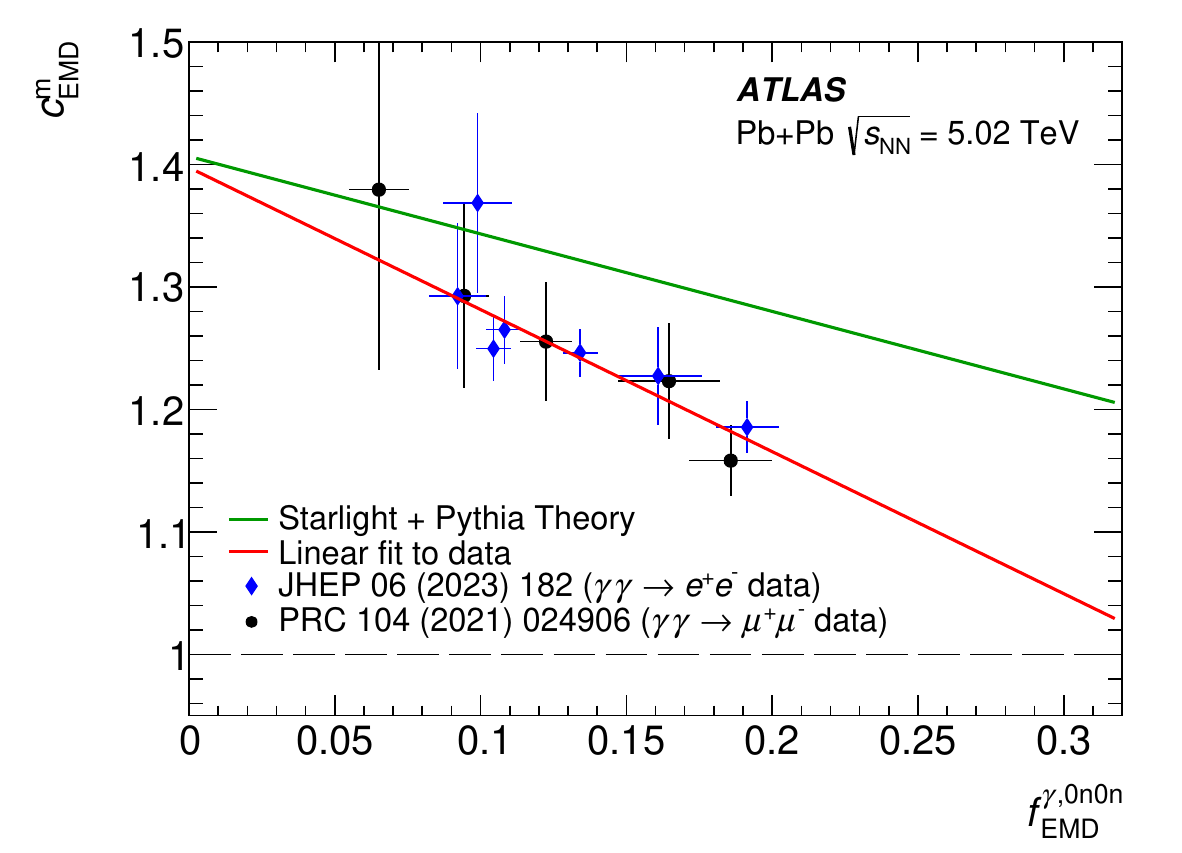}
}
\caption{
Correction factors for electromagnetic dissociation computed through both a data-driven method
and theoretical calculation. The black and blue markers show the binned results
calculated from measurements of $\gamma+\gamma\rightarrow\mu^++\mu^-$ and
$\gamma+\gamma\rightarrow e^++e^-$, respectively. The lines represent combined statistical and
systematic uncertainties. The red line shows the result of a linear fit to this data.
The green line shows the result of a theoretical calculation
for this correction factor which uses \textsc{Starlight}~\cite{Klein:2016yzr} to compute the
probability for EMD as a function of $b_\mathrm{AA}$.
}
\label{fig:CorrEMD}
\end{figure}

Here, \fBU\ is equivalent to the measured quantity derived in Equation~\ref{eq:fnoBU}.
The variables \fBUznxn\ and \fBUznzn\ are the un-corrected equivalent
quantities to \fBU\ as measured in \znxn\ or \znzn\ \gamAjets\
collisions, respectively. In the \znzn\ case, \fBU\ and \fBUznzn\ are the same ($\fBU = \fBUznzn$).
Although \cCorr\ is determined as a function of \fBUznzn\ in each
bin of \zneg, the correction factors are applied to \fBUznxn\ measurements from Ref.~\cite{HION-2022-15}.
In order to derive the correction factor, \cCorr, one may calculate the single and mutual EMD rates in
$\gamma+\gamma\rightarrow\ell^++\ell^-$ measurements, in order to relate them to the corresponding rates
in \gamA\ collisions. The relationship between the rates is found to be:

\begin{equation}
\fBUznzn = \frac{\fznxn}{2 - \fznxn - 2\fxnxn},
\end{equation}

where \fBUznzn\ is the equivalent EMD rate measured for \znzn\ \gamA\ processes, and the quantities
\fznxn\ and \fxnxn\ are the fractions of $\gamma+\gamma\rightarrow\ell^++\ell^-$ collisions which are
measured in that corresponding ZDC topology. Using this relationship, the correction factor, \cCorr,
which applies to \gamA\ interactions is derived according to:
\begin{equation}
\cCorr(\fBUznzn) = \frac{\fznxn + 2\fxnxn}{2\fBUznzn}.
\end{equation}
Since the different
neutron topology fractions are measured for different dilepton kinematics, each set
of kinematics probe a slightly different $b_\text{AA}$
distribution. The available data cover a range in \fBUznzn\ from
about $0.05$ to $0.2$. Results for $\cCorr(\fBUznzn)$ are shown in
Figure~\ref{fig:CorrEMD}. The dependence of \cCorr on \fBUznzn is
approximately linear, and the results of a linear fit are shown in the
figure. That fit is used to provide the correction factors applied to
the \znxn\ data.

In the second approach to estimating \cCorr, theoretical calculations are performed using \textsc{Starlight}~\cite{Klein:2016yzr} to
compute the EMD probabilities as a function of $b_\text{AA}$. Then, by integrating over the
two-nucleus impact parameter, accounting for the nuclear thickness, and multiplying by the EMD probability corresponding to a given
neutron topology, the effective photon flux for each EMD case is computed. These photon fluxes
are then used to re-weight the nominal \pythia\ MC sample and derive \cCorr\ directly, by computing
\fBU\ in both the \znzn\ and \znxn\ cases, then taking the ratio. These correction factors
are also highly linear in \fBUznzn, and they are compared to the separate results from
the above procedure in Figure~\ref{fig:CorrEMD}. The two methods yield
different results for $\cCorr(\fBUznzn)$, reflecting
different underlying assumptions and sensitivity to theoretical
models. The correction factor used for this analysis is taken to be the average of the two
estimates, and an uncertainty, corresponding to half the difference
between the two calculations, is applied. The impact of this correction is
to decrease the corrected \fBU\ from the measured \fBUznxn.


\FloatBarrier
\clearpage

\printbibliography
\clearpage
 
\begin{flushleft}
\hypersetup{urlcolor=black}
{\Large The ATLAS Collaboration}

\bigskip

\AtlasOrcid[0000-0002-6665-4934]{G.~Aad}$^\textrm{\scriptsize 102}$,
\AtlasOrcid[0000-0001-7616-1554]{E.~Aakvaag}$^\textrm{\scriptsize 17}$,
\AtlasOrcid[0000-0002-5888-2734]{B.~Abbott}$^\textrm{\scriptsize 121}$,
\AtlasOrcid[0000-0002-0287-5869]{S.~Abdelhameed}$^\textrm{\scriptsize 83b}$,
\AtlasOrcid[0000-0002-1002-1652]{K.~Abeling}$^\textrm{\scriptsize 54}$,
\AtlasOrcid[0000-0001-5763-2760]{N.J.~Abicht}$^\textrm{\scriptsize 48}$,
\AtlasOrcid[0000-0002-8496-9294]{S.H.~Abidi}$^\textrm{\scriptsize 30}$,
\AtlasOrcid[0009-0003-6578-220X]{M.~Aboelela}$^\textrm{\scriptsize 44}$,
\AtlasOrcid[0000-0002-9987-2292]{A.~Aboulhorma}$^\textrm{\scriptsize 36e}$,
\AtlasOrcid[0000-0001-5329-6640]{H.~Abramowicz}$^\textrm{\scriptsize 154}$
% ...
\bigskip
\\

% ...
$^{17}$Department for Physics and Technology, University of Bergen, Bergen; Norway.\\
% ...
$^{30}$Physics Department, Brookhaven National Laboratory, Upton NY; United States of America.\\
% ...
$^{36}$$^{(a)}$Facult\'e des Sciences Ain Chock, Universit\'e Hassan II de Casablanca;$^{(b)}$Facult\'{e} des Sciences, Universit\'{e} Ibn-Tofail, K\'{e}nitra;$^{(c)}$Facult\'e des Sciences Semlalia, Universit\'e Cadi Ayyad, LPHEA-Marrakech;$^{(d)}$LPMR, Facult\'e des Sciences, Universit\'e Mohamed Premier, Oujda;$^{(e)}$Facult\'e des sciences, Universit\'e Mohammed V, Rabat;$^{(f)}$Institute of Applied Physics, Mohammed VI Polytechnic University, Ben Guerir; Morocco.\\
% ...
$^{44}$Physics Department, Southern Methodist University, Dallas TX; United States of America.\\
% ...
$^{48}$Fakult\"{a}t Physik, Technische Universit{\"a}t Dortmund, Dortmund; Germany.\\
% ...
$^{54}$II. Physikalisches Institut, Georg-August-Universit\"{a}t G\"ottingen, G\"ottingen; Germany.\\
% ...
$^{83}$$^{(a)}$Khalifa University of Science and Technology, Abu Dhabi;$^{(b)}$New York University Abu Dhabi, Abu Dhabi;$^{(c)}$United Arab Emirates University, Al Ain;$^{(d)}$University of Sharjah, Sharjah; United Arab Emirates.\\
% ...
$^{102}$CPPM, Aix-Marseille Universit\'e, CNRS/IN2P3, Marseille; France.\\
% ...
$^{121}$Homer L. Dodge Department of Physics and Astronomy, University of Oklahoma, Norman OK; United States of America.\\
% ...
$^{154}$Raymond and Beverly Sackler School of Physics and Astronomy, Tel Aviv University, Tel Aviv; Israel.\\
% ...
$^{a}$ Also at Affiliated with an institute formerly covered by a cooperation agreement with CERN.\\
$^{b}$ Also at An-Najah National University, Nablus; Palestine.\\
$^{c}$ Also at Borough of Manhattan Community College, City University of New York, New York NY; United States of America.\\
$^{d}$ Also at Center for Interdisciplinary Research and Innovation (CIRI-AUTH), Thessaloniki; Greece.\\
$^{e}$ Also at Centre of Physics of the Universities of Minho and Porto (CF-UM-UP); Portugal.\\
$^{f}$ Also at CERN, Geneva; Switzerland.\\
% ...

\end{flushleft}


%

%
%

%

%
%
%
%

\end{document}